\documentclass[11pt]{article}

\usepackage[latin1]{inputenc}

 \usepackage{amsmath}
 \usepackage{amsfonts}

\usepackage{enumitem}
\usepackage{graphicx}

\usepackage{epstopdf}% To incorporate .eps illustrations using PDFLaTeX, etc.
\usepackage[caption=false]{subfig}% Support for small, `sub' figures and tables

\newcommand{\logit}{\text{logit}}
\newcommand{\var}{\text{var}}
\newcommand{\argmax}{\text{argmax}}
\renewcommand{\P}{\mathbb P}
\newcommand{\E}{\mathbb E}
\usepackage{float} 
\usepackage{enumitem}
\usepackage{dsfont}
\begin{document}

% WORD COUNT

% Words in text: 5253
% Words in headers: 73
% Words outside text (captions, etc.): 300
% Number of headers: 24
% Number of floats/tables/figures: 11
% Number of math inlines: 367
% Number of math displayed: 10

\begin{center}
{\Large
	{\sc  The revisited knockoffs method for variable selection in $L_1$-penalised regressions.}
}
\bigskip

  Anne G\'egout-Petit$^{a}$, Aur\'elie Gueudin-Muller$^{a}$ \& Cl\'emence Karmann$^{a}$
\bigskip

{\it
$^{a}$ Universit\'e de Lorraine, CNRS, Inria, IECL, F-54000 Nancy, France, Inria BIGS Team ;  anne.gegout-petit@univ-lorraine.fr, aurelie.gueudin@univ-lorraine.fr, clemence.karmann@inria.fr
}
\end{center}
\bigskip

{\bf Abstract.} We consider the problem of variable selection in regression models. In particular, we are interested in selecting explanatory covariates linked with the response variable and we want to determine which covariates are relevant, that is which covariates are involved in the model. In this framework, we deal with $L_1$-penalised regression models. To handle the choice of the penalty parameter to perform variable selection, we develop a new method based on the knockoffs idea. This revisited knockoffs method is general, suitable for a wide range of regressions with various types of response variables. Besides, it also works when the number of observations is smaller than the number of covariates and gives an order of importance of the covariates. Finally, we provide many experimental results to corroborate our method and compare it with other variable selection methods.
\smallskip


%---------------------------------------------------------------------------------------

\section{Introduction}

Regression methods are really helpful to analyse dependencies between a variable, named the response, and one or several explanatory covariates. This is one of the reasons why they are widely used and studied in statistical analysis \cite{BookHastieTibshFriedman}. Many models have been introduced %to respond to natural demands
including the well-known linear regression for a continuous response variable or logistic regression for a binary response variable. Indeed, many data sets involve this last situation such as the occurence of a disease in medicine or voting intentions in econometrics. Another type of data is nominal data (that is unordered categorical data) like housing types or food choice of predators. The situation is a bit more complicated when the response is ordered categorical (ordinal), e.g. different stages of cancer, pain scales, place ratings on Google or data collected from surveys (0: never, 10: always). Logistic regression can naturally be extended to the case where the response is nominal. For such data, many authors \cite{BookAgrestiOrdinal,McCullagh1980, LiuAgresti2005,Suggala2017} provided models based on odds ratios such as cumulative link models, adjacent-categories logit models or continuation-ratio logit models. The choice of one of these models depends on the kind of problem. In this paper, we concentrate on a restricted but large spectrum of regression models including all regression models mentioned above.\\

Although prediction and interpretation provide major challenges in regression motivations, another important issue is to identify the influential explanatory covariates, that is variable selection. Selection problems often arise in many fields including biology \cite{WhuChenHastie}. For example, in microarray cancer diagnosis \cite{ZhuHastie2004}, a primary goal is to understand which genes are relevant. For cost and time reasons, it can also be convenient for biologists to restrict their studies to a smaller subset of explanatory covariates (genes, bacteria populations...). Accordingly, the sparsity assumption (that is, a few number of relevant explanatory covariates) is frequently suitable and adequate, even crucial for interpretation. Indeed, with a large number of covariates, it is also useful for interpretation to determine a smaller subset of covariates that have the strongest effects. Besides, when the number of covariates is larger than the number of observations or when covariates are highly correlated, standard regression methods become inappropriate.
 
Lasso penalisation, or $L_1$-penalisation, introduced by Tibshirani \cite{Tibshirani1997} offers an attractive solution to these issues. That includes a $L_1$ penalty in the estimation of the regression coefficients in order to perform variable selection by optimising a convex criterion. The regularisation resulting from Lasso penalisation shrinks down to zero the coefficients of explanatory covariates that have the lowest effects and leads to sparse solutions and more interpretable models, making Lasso one of the most popular penalisation \cite{BookHastie2015StatisticalSparsity,ZhaoYu2006,ParkHastie2007}. %However, obtaining such models sometimes involves heavy optimisation issues.
\\

Moreover, using Lasso induces the critical choice of the penalty parameter which controls the number of selected covariables. This choice is major because two close values of the penalty parameter can often lead to very different scientific conclusions. Many general techniques have been proposed in the literature but they do not have the same purposes. For instance, $K$-fold cross validation emphasises prediction, the validation step involving computing the prediction error and aiming at minimising this. Furthermore, cross validation is often quite greedy and tends to overfit the data \cite{Wasserman2009high}. Other techniques, like StARS \cite{StARS}, can be adapted to a regression framework and aim at 'overselecting', that is selecting a larger set of covariates which contains the relevant ones, allowing false positive detections. Some frameworks such as gene regulatory networks require this choice: indeed, false positive detections can then be eliminated by further biological experiments whereas omitted interactions cannot be recovered after that. On the contrary, one can prefer selecting a set of covariates included in the set of true covariates to avoid false positive detections that is 'underselecting'. This constraint comes from the fact that after selection, the relevant covariates have to be studied by scientists through new experiments. But new experiments are generally expensive or time-consuming and it would be a waste to involve noisy covariates. In this paper, we focus on variable selection in the former case. Compared with our goals, we propose an intuitive and general method for automatic variable selection, inspired from the knockoffs idea of Barber and Candes \cite{CandesKnockoff,Candes2016knockoff}. This method uses a matrix of knockoffs of the covariates built by swapping the rows of the matrix of covariates. This knockoffs matrix is thus random and aims at determining if a covariable belongs to the model using a decision rule based on change detection methods. One of the major advantages is that it can be performed in a wide range of regression frameworks including when the number of covariates is much larger than the number of observations. We will see that our method does not lead to a choice of the penalty parameter. Nevertheless, it provides an order of importance on the covariates allowing to select covariates according to the target.\\

In this paper, we address the problem of variable selection in $L_1$-penalised regressions. Our goal is to determine which covariates are relevant and which are noisy. We achieve it by proposing a new method of type knockoffs. The rest of the paper is organised as follows. In Section 2, we first introduce the background and describe the knockoffs method for variable selection. We also describe briefly our R package \texttt{kosel} in which the revisited knockoffs method is implemented. In Section 3, we give many illustrations and results of our method on simulated data. Furthermore, we propose a way to exploit randomness of our procedure.

%---------------------------------------------------------------------------------------

% materials and methods

\section{Revisited knockoffs method}

\subsection{Background}

Consider we have $p$ explanatory $\mathbb R$-valued covariates $\overrightarrow{X} := (X_1, X_2, \dots, X_p)$ and a response variable $Y$ linked with $\overrightarrow{X}$ by $m$ equations of the type:
\begin{equation}\label{regression}
f_k(\mu_k(Y|X)) = \alpha_k + \beta_1 X_1 + \ldots + \beta_p X_p,\ k = 1, \ldots, m,
\end{equation}
where $f_k$ is a deterministic function, $\mu_k(Y|X)$ parameters of the distribution of $Y$ given $X$ and $\alpha_k, \beta_1, \ldots, \beta_p$ real coefficients. Note that the vector of regression coefficients $\boldsymbol \beta := (\beta_1, \ldots, \beta_p)$ does not depend on $k$.\\

This framework is quite general and emcompasses many regression models such as generalised linear models \cite{BookAgrestiCategorical} (linear regression, logistic regression, Poisson regression, multinomial regression), ordinal logistic regression models \cite{BookAgrestiOrdinal} (cumulative logit models with proportional odds \cite{Simon1974,WilliamsGrizzle1972,AndersonPhilips1981}, adjacent-categories logit models, continuation-ratio logit models) or cumulative link models \cite{BookAgrestiOrdinal}. Indeed, for generalised linear models, $m = 1$, $\mu_1(Y|X) = \E(Y|X)$ and $f_1$ is the link function (identity, $\log$, $\logit$...) of the corresponding model. For ordinal logistic regression models, $f_k = \logit$ and $\mu_k(Y|X) = \P(Y \leq k | X)$ (cumulative), $\mu_k(Y|X) = \P(Y = k | Y = k\text{ or } Y = k + 1, X)$ (adjacent), $\mu_k(Y|X) = \P(Y > k | Y \leq k, X)$ (continuation). Notice that these last models only allow identical effects of the covariates, which implies that the regression coefficients $\beta_i$ do not depend on the modality $k$ of the response variable $Y$. In particular, this framework includes models for many types of response variable such as binary, continuous, ordinal or categorical.\\

In this framework, covariates $X_l,\ l = 1, \ldots, p$ have to be linked to the response variable $Y$ through a linear expression so that the conditional dependence between $Y$ and $X_l$ given $X_1, \ldots, X_{l-1}, X_{l+1}, \ldots, X_p$ can be measured through the regression coefficient $\beta_l$. More precisely, $\beta_l = 0$ means that $X_l$ and $Y$ are independent conditionally on the other covariates $X_k,\ k = 1, \ldots, l-1, l+1, \ldots p.$ We are thus interested in the nullity of the regression coefficients $\beta_l$ to select the relevant covariates. Moreover, we make sparsity assumption, that is a relatively small number of covariates play an important role. This implies that only a few covariates are relevant and thus, only a few regression coefficients $\beta_l$ are non-null. This sparsity assumption is convenient for scientists to restrict their studies to a smaller subset of covariates, namely in high-dimensional settings. Instead of checking the nullity of each coefficient $\beta_l$ by performing statistical tests, we add a $L_1$-penalisation on the coefficients $\boldsymbol{\beta}$ in the estimation of the coefficients of the model. Coefficients are usually estimated by solving the optimisation problem:
\begin{equation}\label{EstiClassique}
\underset{(\boldsymbol{\alpha},\boldsymbol{\beta})}{\argmax\ } L(\boldsymbol{\alpha}, \boldsymbol{\beta},\textbf{Y},\textbf{X}),
\end{equation}
where $L(\boldsymbol{\alpha}, \boldsymbol{\beta}, \textbf{Y}, \textbf{X})$ is a function of the coefficients relative to the model (like log-likelihood), depending on the observations $\textbf{Y}$ and $\textbf{X}$ of the response variable $Y$ and the vector of covariates $\overrightarrow{X}$ respectively. Instead of estimating the coefficients by \eqref{EstiClassique}, we add a Lasso penalisation on the coefficients vector $\boldsymbol{\beta}$ which leads to solve the following optimisation problem:
\begin{equation}\label{penaloptim}
\underset{(\boldsymbol{\alpha},\boldsymbol{\beta})}{\argmax\ } \big\{L(\boldsymbol{\alpha}, \boldsymbol{\beta}, \textbf{Y}, \textbf{X}) - \lambda ||\boldsymbol{\beta}||_1\big\},
\end{equation}
where $\lambda > 0$ is the penalty parameter. \\

Usually, all penalisation methods require the choice of the (positive) penalty parameter, also referred as tuning or regularisation parameter. We then need to tune the penalty parameter $\lambda$ (involved in the optimisation problem \eqref{penaloptim}) which controls the number of selected covariates: the larger $\lambda$ is, the fewer the selected covariates are. On the contrary, values of $\lambda$ closed to $0$ lead to the full model, that is the model with all the covariates. We remind that our goal is to select only relevant covariates and thus, to avoid false positive detections (the wrongly detected covariates). \\

With regard to our problems and goals, we propose a new method, inspired from the knockoffs process used by Barber and Candes \cite{CandesKnockoff} in the linear Gaussian regression setting. Actually, this method does not lead to a choice of the penalty parameter $\lambda$ but it puts the explanatory covariates in order from the most relevant to the least%, allowing the user to make its own choice
. Furthermore, it suits any regression of the type presented in \eqref{regression} including when the number $n$ of observations is smaller than the number $p$ of covariates. Obviously, in the linear Gaussian model, it is much more pertinent to use the procedure described in \cite{CandesKnockoff} because of their theoretical guarantees. Even if their procedure initially held for $n > p$, they subsequently extended it thanks to a preliminary screening step \cite{Candes2016knockoff}. In what follows, we present the principle of our revisited knockoffs method.\\

\subsection{Principle and generalities}

Let $\textbf{X}$ denote the $n\times p$ matrix of the $n$ observations of the $p$-vector $\overrightarrow{X} = (X_1,\ldots,X_p)$ of covariates, called the design matrix.
The principle, given in \cite{CandesKnockoff}, is to use a matrix $\tilde{\textbf{X}}$ of knockoffs (of the covariates $X_i$) whose covariance structure is similar to $\textbf{X}$ but independent from $\textbf{Y}$. The goal is to determine if a covariate $X_i$ is relevant by studying if it enters the model before its knockoff $\tilde X_i$, that is if $X_i$ enters the model for a larger value of the penalty parameter $\lambda$. Indeed, as the knockoff matrix is independent from $\textbf{Y}$, if a covariate enters the model after its knockoff, we can rightfully suspect that this covariate does not belong to the model.\\

We mainly differ from the method proposed by \cite{CandesKnockoff} in the construction of the knockoffs. In their paper, they propose a sophisticated construction of the knockoff filter using linear algebra tools. This construction allows to control the false discovery rate (FDR) --the expected fraction of false discoveries among all discoveries-- in the linear Gaussian model whenever there are at least as many observations as covariates. This difference in the construction of the knockoffs makes our method suitable for the setting $n < p$ and for a larger set of regression models. Nevertheless, theoretical guarantees about the control of the false discovery rate do not hold anymore.\\ 
We construct our knockoff matrix $\tilde{\textbf{X}}$ by randomly swapping the $n$ rows of the design matrix $\textbf{X}$. This way, the correlations between the knockoffs remain the same as the original variables but the knockoffs are not linked to the response $\textbf{Y}$. Note that this construction of the knockoffs matrix also makes the procedure random. Then, in the same way as \cite{CandesKnockoff}, we perform the regression of $\textbf{Y}$ on the $n \times 2p$ augmented matrix $[\textbf{X}, \tilde{\textbf{X}}]$, i.e. the columnwise concatenation of $\textbf{X}$ and $\tilde{\textbf{X}}$. Let us note $\boldsymbol{\hat\beta}(\lambda)$ the estimated regression coefficients of the $\lambda$-penalised regression of $\textbf{Y}$ on the augmented matrix $[\textbf{X}, \tilde{\textbf{X}}]$:
\begin{equation*}
\Big(\boldsymbol{\hat\alpha}(\lambda),\boldsymbol{\hat\beta}(\lambda)\Big) := \underset{(\boldsymbol{\alpha},\boldsymbol{\beta})}{\argmax\ } \big\{L(\boldsymbol{\alpha}, \boldsymbol{\beta}, \textbf{Y}, [\textbf{X}, \tilde{\textbf{X}}]) - \lambda ||\boldsymbol{\beta}||_1\big\}.
\end{equation*} 
For each variable of the augmented design, that is for each covariate and its corresponding knockoff, we consider $T_i := \sup\ \{\lambda>0,\ \hat\beta_i(\lambda) \neq 0\},\ i \in \{1,\ldots,p,p+1,\ldots,2p\}$. Statistics $T_i$ correspond to the largest value of $\lambda$ for which the covariate $X_i$ if $i \in \{1,\ldots,p\}$ or its knockoff $\tilde X_{i-p}$ if $i \in \{p+1,\ldots,2p\}$ first enters the model. At this stage, we hope that $T_i$ is large for the relevant covariate, that is for $X_i, i \in \{1,\ldots,p\}$ such that $\beta_i \neq 0$ and small for the knockoffs variables $X_i := \tilde X_{i-p},\ i \in \{p+1,\ldots,2p\}$ or for the noisy covariates $X_i,\ i \in \{1,\ldots,p\}$ such that $\beta_i = 0$. This yields us a $2p$-vector $(T_1,...,T_p,\tilde T_1,...,\tilde T_p)$ where $\tilde T_i$ denotes $T_{i+p}$. Then, we consider, for all $i \in \{1,...,p\}$, $W_i := \max(T_i, \tilde T_i) \times \left\{
    \begin{array}{ll}
        (+1) & \mbox{if } T_i > \tilde T_i \\
        (-1) & \mbox{if } T_i \leq \tilde T_i
    \end{array}
\right.$. 

Statistics $W_i$ allo to determine if a covariate enters the model before or after its knockoff. A negative value for $W_i$ means that the covariate $X_i$ enters the model after its knockoff and we eliminate it. On the contrary, a positive value for $W_i$ means that the covariate $X_i$ enters the model before its knockoff and is more likely to belong to the model. But covariates $X_i$ whose statistic $W_i$ is positive are not necessarily relevant: we hope that $W_i$ is large for most of relevant covariates and small for the other ones. 
Thus, we are interested in the largest positive values of the $p$-vector of statistics $W$ which moreover indicates that the corresponding covariate enters the model early, that is for a large value of $\lambda$. Statistics $W_i$ allows in fact to sort the covariates according to their importance: the larger $W_i$ is, the more relevant the associated covariate $X_i$ is.\\ 
This then implies defining a threshold $s$ for $W_i$ over which we will keep the corresponding covariates in the estimated model. On the whole, we will choose the estimated model $\hat S$ such that:
\[ \hat S := \{X_i : W_i \geq s\}.
\]

\subsection{Choice of the threshold} \label{Thresholds}

\begin{figure}
\centering
\includegraphics[scale = 0.55]{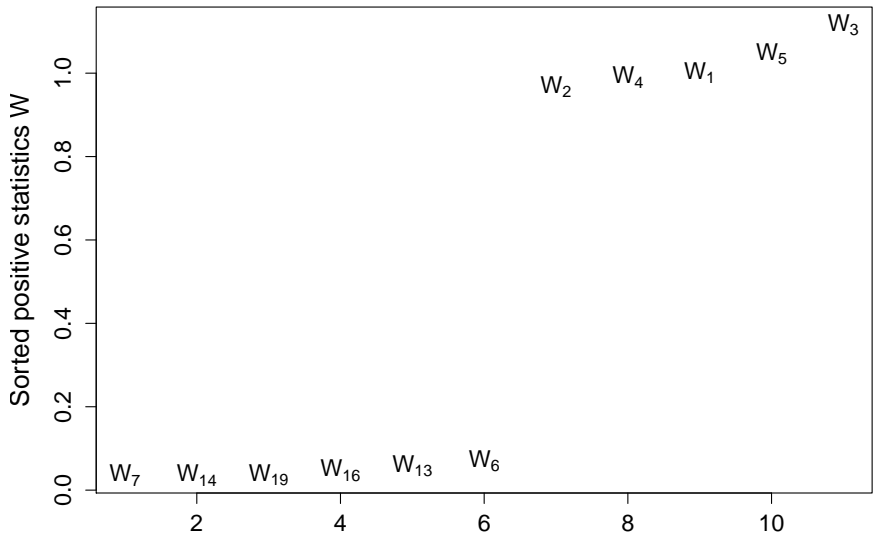}
\caption{Example of positive statistics $W_i$ sorted in ascending order. Linear Gaussian regression model with $n = 500$ observations of $p = 20$ covariates. Only covariates $X_1$, $X_2$, $X_3$, $X_4$ and $X_5$ belong to the model (regression coefficients are set to $\beta = (1,1,1,1,1,0,\ldots,0)$).} \label{Distributions}
\end{figure}

The second major difference with Barber and Candes \cite{CandesKnockoff} lies in the choice of the threshold $s$. They provide in fact a data-dependent threshold that shows attractive results relative to the false discovery rate in the Gaussian setting. Unfortunately, these results do not hold in general, out of the linear Gaussian model. In our method, we make the assumption that there is a breakdown in the distributions between the $W_i$ corresponding to the covariates $X_i$ belonging to the model and the other ones (see Figure \ref{Distributions}). Figure \ref{Distributions} illustrates that distributions of $W_i$ depend on whether $X_i$ is relevant or not. To generate Figure \ref{Distributions}, we have simulated a set of data under a linear Gaussian regression model with $p = 20$ independent Gaussian covariates. Only the five first ones were linked to $Y$:
$$Y = \beta_1X_1 + \dots + \beta_p X_p + \epsilon, $$
where $\beta = (1,1,1,1,1,0,\dots,0)$ and $\epsilon \sim \mathcal N(0,1)$. In our knockoffs procedure applicated to this data set, only statistics $W_1,W_2,W_3,W_4, W_5,W_6, W_7,W_{13},W_{14},W_{16}, W_{19}$ associated to the covariates $X_1, X_2, \ldots, X_{14}, X_{16}$ and $X_{19}$ have a positive value. For example, the covariate $X_1$ entered the model for $\lambda = 1.002$ (thus, $T_1 = 1.002$) and entered the model before its knockoff $\tilde X_1$. This means that the knockoff variable $\tilde X_1$ entered the model for $\lambda < 1.04$ and implies that $\tilde T_1 < 1.002$. $W_3$ takes the largest value among all the statistics $W_i,\ i = 1, \ldots, 20$, which implies that $X_3$ is the covariate the most likely to belong to the model. We can clearly observe a breakdown between the values of the five first covariates and the other ones.\\

Consequently, we present two automatic ways to define the threshold $s$ by using two change detection methods: the method proposed by Auger and Lawrence \cite{AugerLawrence,picard2004statistical,Lebarbier} and the CUSUM method for mean change detection. Let  $W_{(i)},\ i = 1,\ldots,w,$ denote the sorted $w$ positive statistics $W_i,\ i = 1,...,w$, that is $0 < W_{(1)} \leq W_{(2)} \leq \ldots \leq W_{(w)}$ and $e_j~=~W_{(j+1)}~-~W_{(j)}$ for all $j = 1, \ldots w-1$ the $w-1$ gaps between these sorted statistics. Remark that $w$, the number of positive statistics $W_i$, is random ($w = 11$ on Figure \ref{Distributions}). We propose two automatic thresholds defined as:
\begin{itemize}
\item the minimum of the two thresholds obtained by applying these two change detection methods directly on the statistics $W_{(i)},\ i = 1, \ldots, w$ sorted in ascending order,
\item the minimum of the two thresholds obtained by applying these two change detection methods on the gaps $e_j,\ j = 1, \ldots, w-1$.
\end{itemize} 
Let us name the first one '$W$-threshold' and the second one 'gaps-threshold' for the sake of simplicity.

\subsection{R package \texttt{kosel}}

Our procedures have been implemented in a R package, called \texttt{kosel} (for knockoffs selection), available on the CRAN. Our package includes three functions: \texttt{ko.glm}, \texttt{ko.ordinal} and \texttt{ko.sel}.\\

The two first functions construct the knockoffs matrix and return the $p$-vector of statistics $W$ for $L_1$-penalised regressions models respectively implemented in the R functions \texttt{glmnet} and \texttt{ordinalNet} from the packages of the same name. \texttt{glmnet} emcompasses generalised linear models whereas \texttt{ordinalNet} includes ordinal regression models such as cumulative link, adjacent or continuation-ratio or stopping-ratio. By default, a seed is used so that the knockoffs matrix remains the same (and thus, the resulting statistics vector $W$). But this can be changed with the option \texttt{random = TRUE} to exploit the randomness of the procedure (see Subsection \ref{Randomness} for further details).\\

The third function \texttt{ko.sel} deals with the choice of the threshold. It uses the statistics vector $W$ obtained by one of the two other functions and returns the $p$-binary vector of estimation and the threshold $s$. Three choices are proposed: \texttt{method = 'stats'} and \texttt{method = 'gaps'} respectively correspond to the '$W$-threshold' and 'gaps-threshold' while \texttt{method = 'manual'} allows the user to choose its own threshold. The option \texttt{print = TRUE} displays the positive statistics $W_i$ sorted in ascending order like in Figure \ref{Distributions}. For \texttt{method = 'manual'}, they are automatically displayed so that the user can choose its threshold. For the '$W$-threshold' (\texttt{method = 'stats'}) and 'gaps-threshold' (\texttt{method = 'gaps'}), option \texttt{print = TRUE} also displays an horizontal line corresponding to the threshold.

%---------------------------------------------------------------------------------------

% results 

\section{Simulation studies}

\subsection{Settings}

We now describe experimental results to study the efficiency of our procedure. For that, we have performed different simulations with various regressions: linear Gaussian regression, logistic regression and cumulative logit regression (with proportional odds). Covariates $\overrightarrow X$ are simulated as Gaussian such that  $\E(X_k) = 0$ and $\var(X_k) = 1$ for all $k = 1, \ldots, p$ and such that $X_i$ and $X_j$ are dependent conditionally on the other covariates $X_k,\ k \in \{1, \ldots, p\}\setminus\{i,j\}$ with probability $0.2$. The design matrix $\textbf{X}$ of covariates has been simulated with the R function \texttt{huge.generator} from the package \texttt{huge}, for a random graph structure. We have then simulated the observations of the response variable $Y$ as:
\begin{align*}
Y & = \beta_1 X_1 + \ldots + \beta_p X_p + \epsilon, \tag{linear regression}\\
\logit(\P(Y = 1|X)) & = \alpha_1 + \beta_1 X_1 + \ldots + \beta_p X_p, \tag{logistic regression}\\
\text{or } \logit(\P(Y \leq k|X)) & = \alpha_k + \beta_1 X_1 + \ldots + \beta_p X_p, k = 1, 2, \tag{cumulative logit regression}
\end{align*}
where $\epsilon \sim \mathcal N(0, 1)$ is Gaussian noise, the vector of regression coefficients $\boldsymbol \beta$ is sparse and given below and intercepts $\alpha_1$ and $\alpha_2$ are chosen so that the response variable $\textbf{Y}$ takes enough values in each of its modalities ($\{0,1\}$ for logistic regression and $\{0,1,2\}$ for cumulative logit regression). These regressions have been respectively performed with the R functions \texttt{glmnet} and \texttt{ordinalNet} of the eponymous R packages.\\

We present detection rates of each covariates on $B = 100$ repetitions for different settings. The detection rate of the covariate $X_l$ is the number of times among the $100$ repetitions where the estimated model included $X_1$. First, we have simulated $n = 200$ observations of $p = 50$ covariates for pedagogical reasons and next, $n = 1000$ observations of $p = 2000$ covariates to illustrate results in a higher-dimensional framework. For $p = 2000$ covariates, results are presented as boxplots of detection rates according to the regression coefficient $\beta$ in order to improve readability.\\ 
In addition, we compare our results with results obtained by cross validation. Cross validation has been performed with the R functions \texttt{cv.glmnet} for linear and logistic regressions and \texttt{ordinalNetTune} with 'logLik' tune method for cumulative logit regression. For $p = 50$, we also compare our results in the linear Gaussian setting with results obtained by Barber and Candes' procedure. Their procedure is implemented in the function \texttt{knockoff.filter} from the R package \texttt{knockoff}. We do not perform this comparison for $p = 2000$ because their procedure is not applicable due to the too few observations.

\subsection{Efficiency and comparisons}

\subsubsection{p = 50} \label{p50}

In the first place, we present results for $n = 200$ observations of $p = 50$ covariates. Regression coefficients are set to $\boldsymbol \beta = (1,1,1,1,1,0,\ldots,0)$ and $\boldsymbol \beta = (2.5,2,1.5,1,0.5,0,\ldots,0)$. Covariates $\textbf{X}$ are the same for each different regression. But they are different according to the regression coefficients $\boldsymbol \beta$. In other words, for a fixed value of $\boldsymbol \beta$, the design matrix $\textbf{X}$ is the same for each of the three regression models. But the response variable $\textbf{Y}$ is simulated according to the regression model and is therefore different. The knockoffs matrix is also different.\\

\begin{figure}[htbp!]
\centering
\subfloat[$\boldsymbol \beta = (1,1,1,1,1,0,\ldots, 0)$.]{%
\resizebox*{7.11cm}{!}{\includegraphics{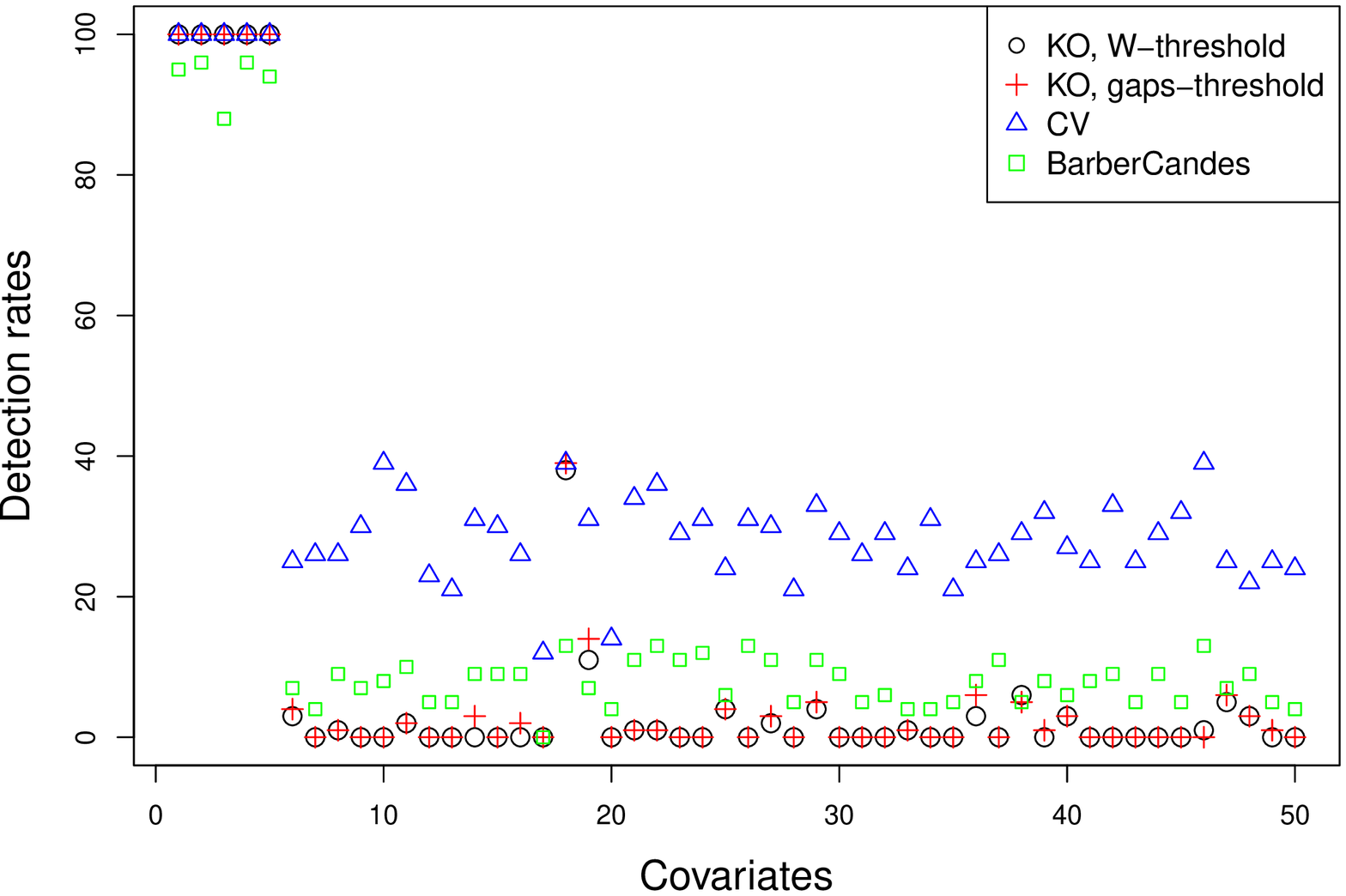}}}\hspace{0pt}
\subfloat[$\boldsymbol \beta = (2.5,2,1.5,1,0.5,0,\ldots,0)$.]{%
\resizebox*{7.11cm}{!}{\includegraphics{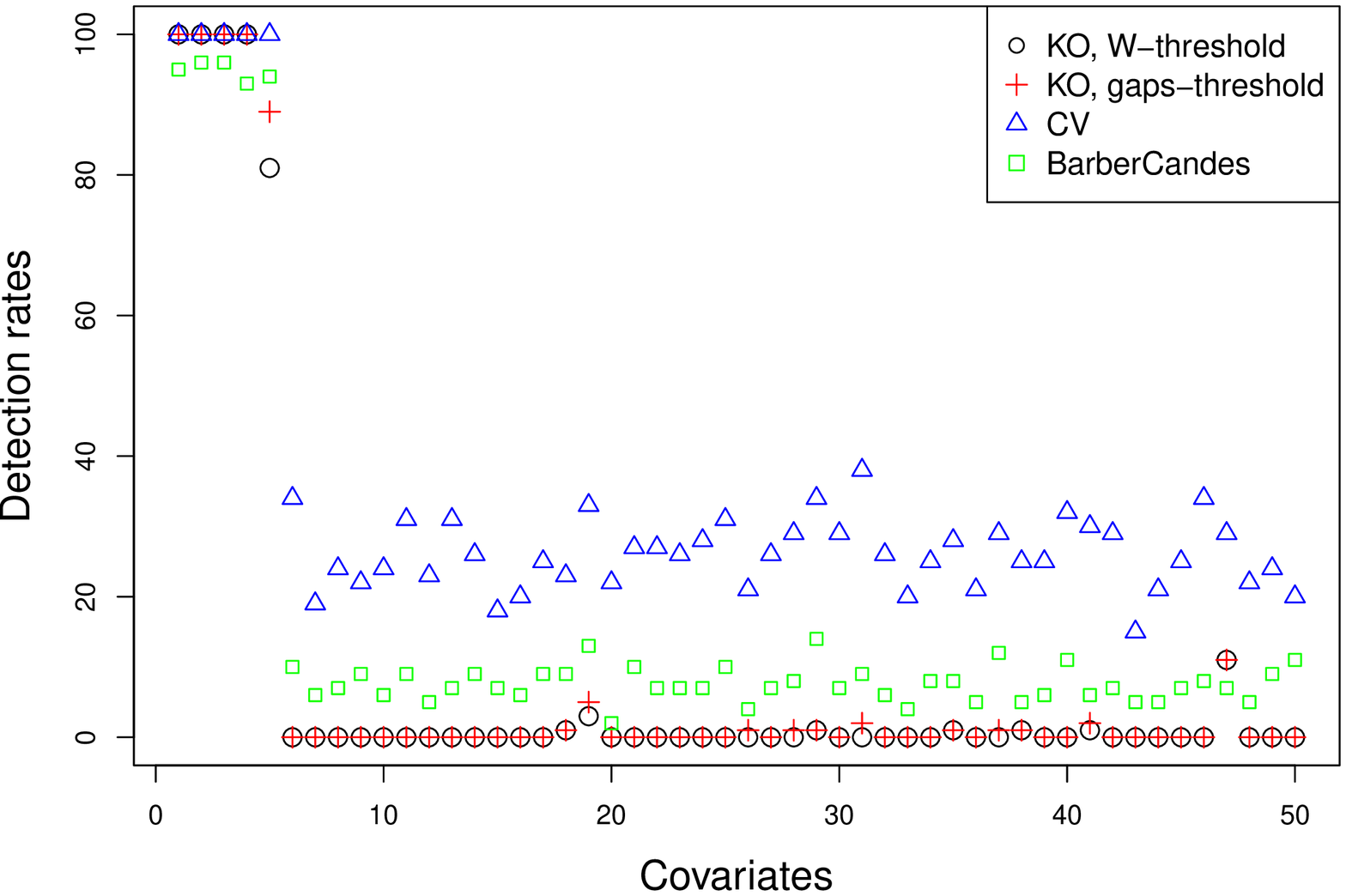}}}
\caption{Detection rates of each covariate for the four methods: revisited knockoffs $W$-threshold and gaps-thresholds, cross validation and Barber and Candes' knockoffs. Linear Gaussian regression model with $n = 200$ observations of $p = 50$ covariates with regression coefficients $\boldsymbol \beta = (1,1,1,1,1,0,\ldots, 0)$ (a) and $\boldsymbol \beta = (2.5,2,1.5,1,0.5,0,\ldots,0)$ (b). Covariates are dependent Gaussian with a random structure. The number of i.i.d. repetitions is $B = 100$.} \label{fig:RLG_EC}
\end{figure}

\begin{figure}[]
\centering
\subfloat[$\boldsymbol \beta = (1,1,1,1,1,0,\ldots, 0)$.]{%
\resizebox*{7.11cm}{!}{\includegraphics{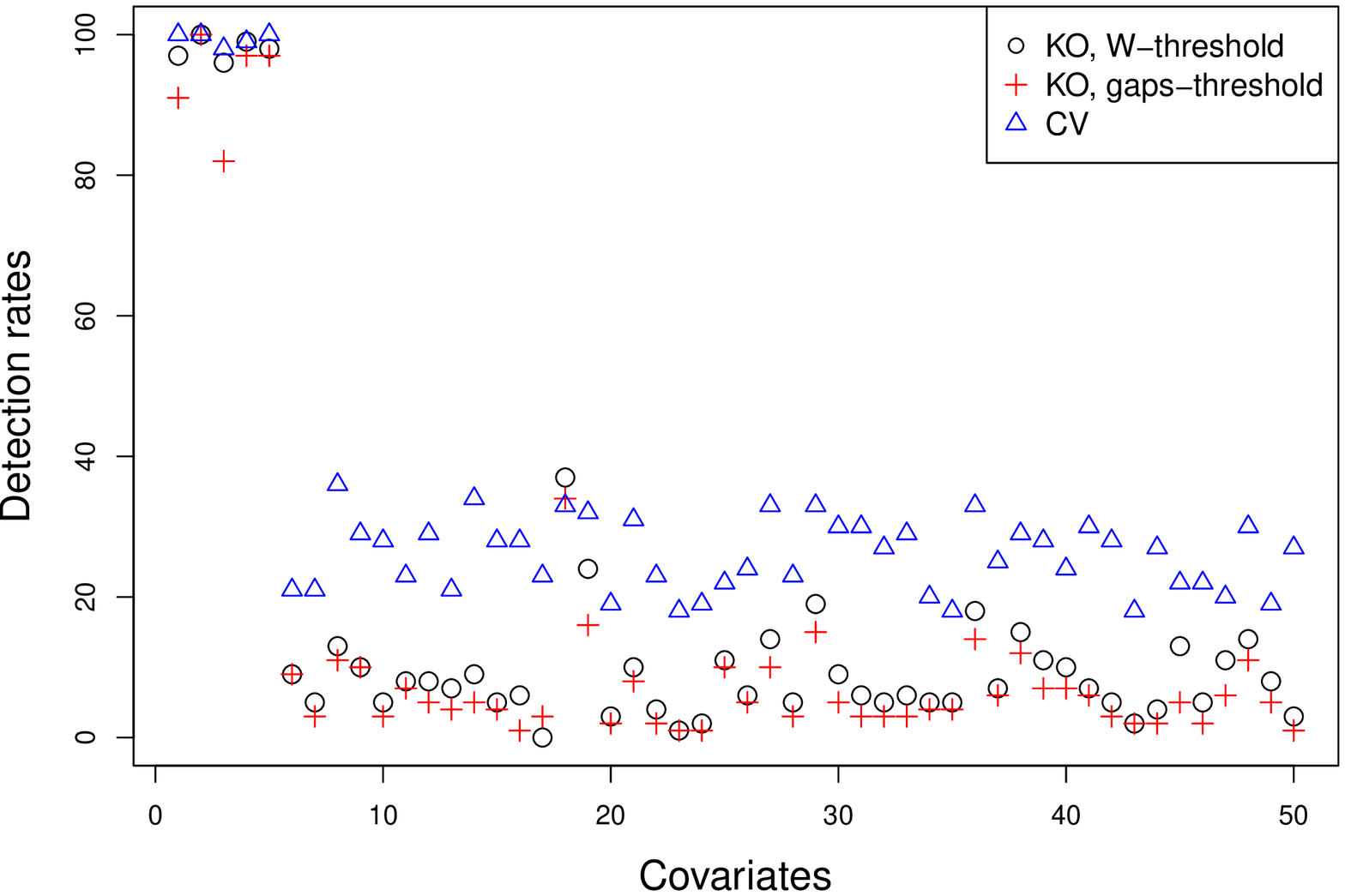}}}\hspace{0pt}
\subfloat[$\boldsymbol \beta = (2.5,2,1.5,1,0.5,0,\ldots,0)$.]{%
\resizebox*{7.11cm}{!}{\includegraphics{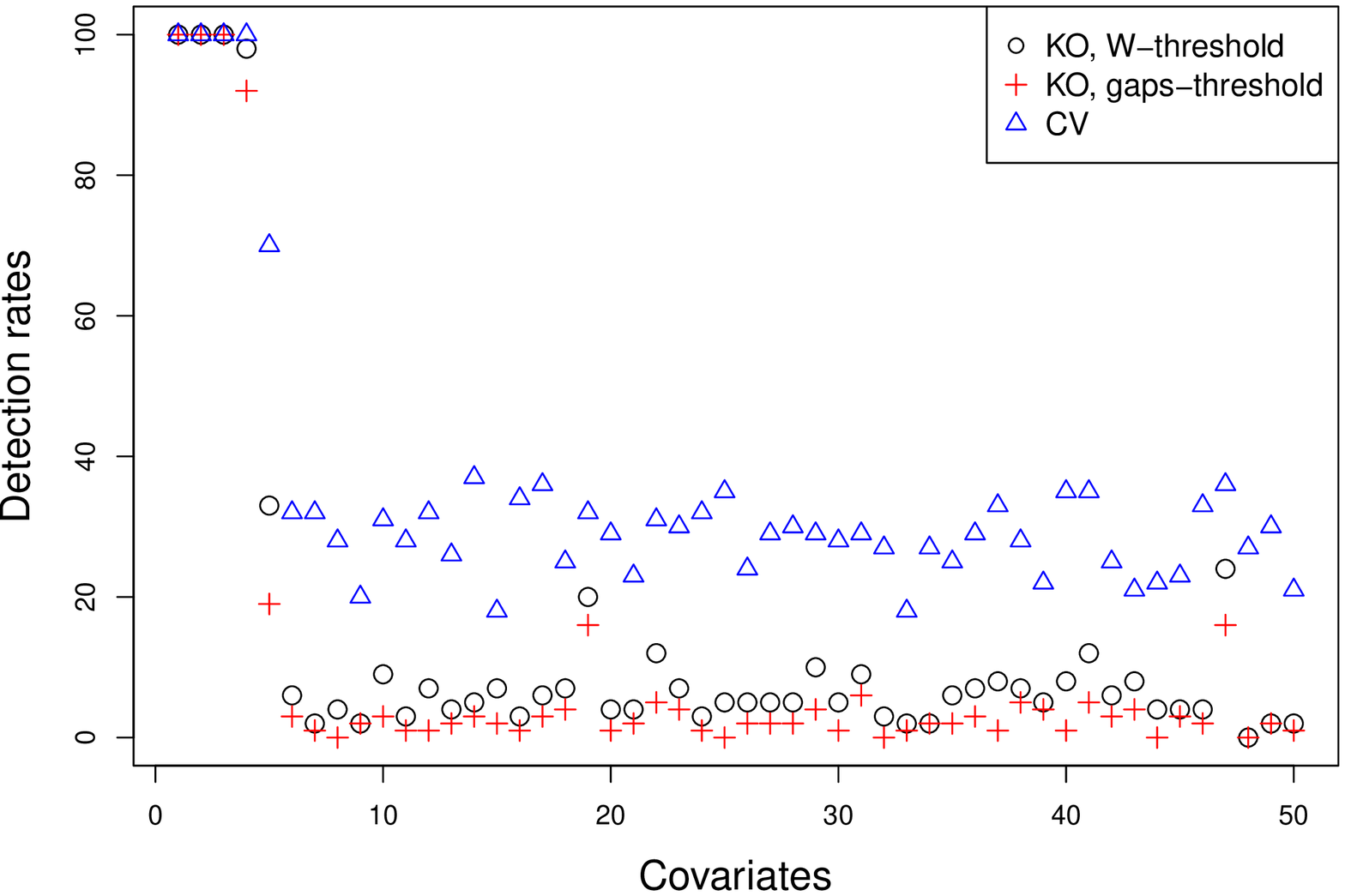}}}
\caption{Detection rates of each covariate for the three methods: revisited knockoffs $W$-threshold and gaps-thresholds and cross validation. Logistic regression model with $n = 200$ observations of $p = 50$ covariates with regression coefficients $\boldsymbol \beta = (1,1,1,1,1,0,\ldots, 0)$ (a) and $\boldsymbol \beta = (2.5,2,1.5,1,0.5,0,\ldots,0)$ (b). Covariates are dependent Gaussian with a random structure. The number of i.i.d. repetitions is $B = 100$.} \label{fig:RL_EC}
\end{figure}

\begin{figure}[]
\centering
\subfloat[$\boldsymbol \beta = (1,1,1,1,1,0,\ldots, 0)$.]{%
\resizebox*{7.11cm}{!}{\includegraphics{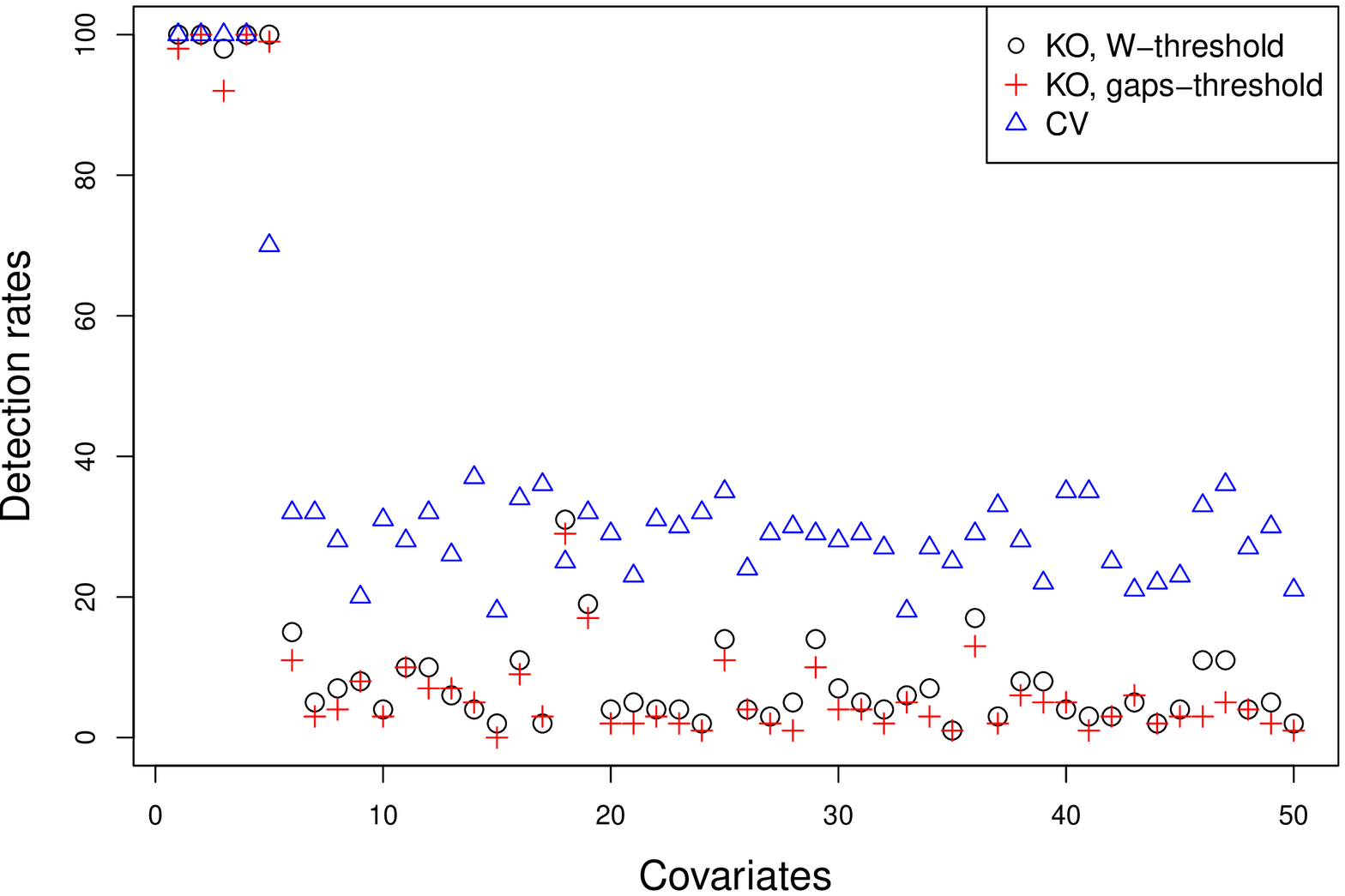}}}\hspace{0pt}
\subfloat[$\boldsymbol \beta = (2.5,2,1.5,1,0.5,0,\ldots,0)$.]{%
\resizebox*{7.11cm}{!}{\includegraphics{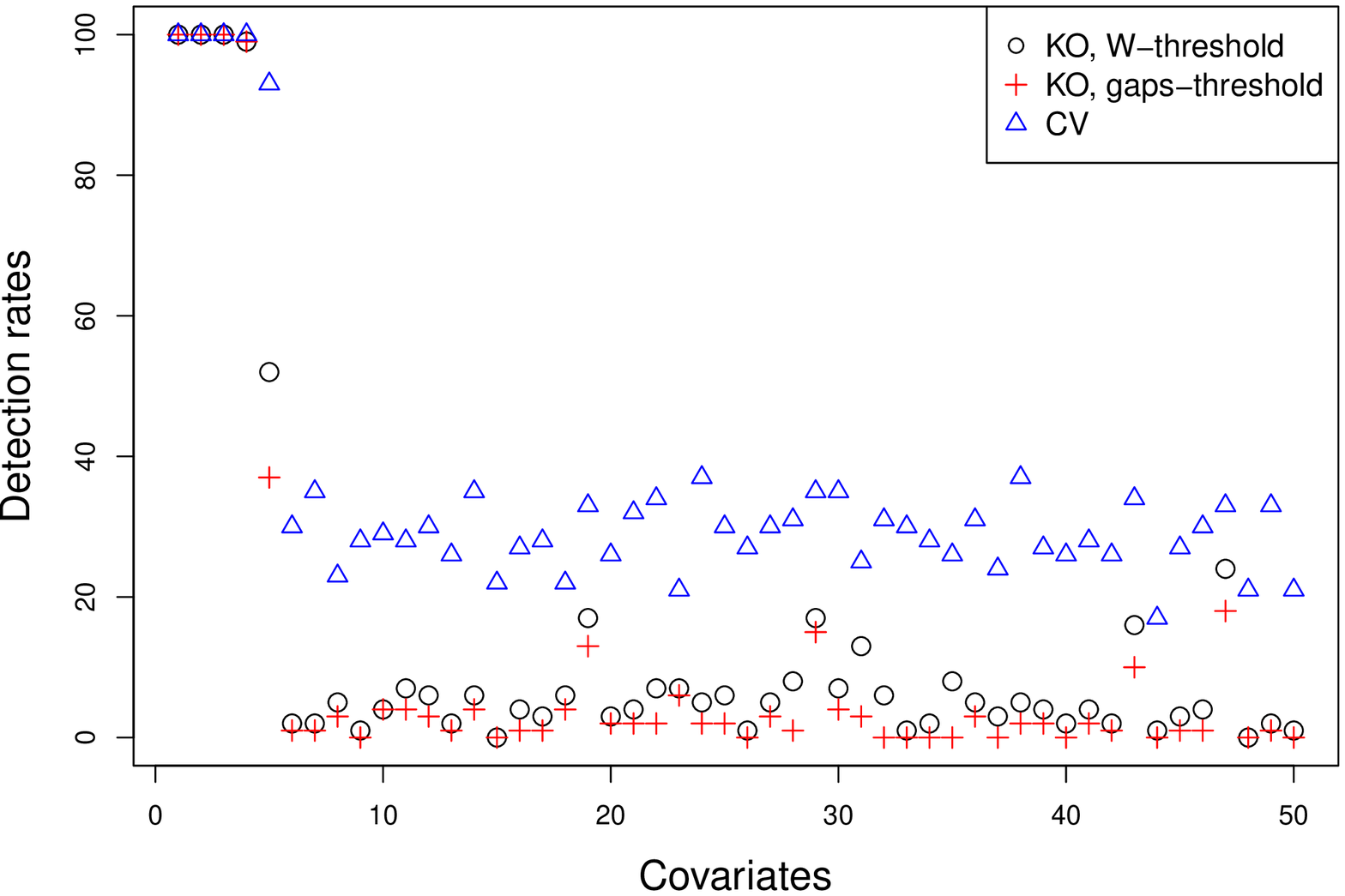}}}
\caption{Detection rates of each covariate for the three methods: revisited knockoffs $W$-threshold and gaps-thresholds and cross validation. Cumulative logit model with $n = 200$ observations of $p = 50$ covariates with regression coefficients $\boldsymbol \beta = (1,1,1,1,1,0,\ldots, 0)$ (a) and $\boldsymbol \beta = (2.5,2,1.5,1,0.5,0,\ldots,0)$ (b). Covariates are dependent Gaussian with a random structure. The number of i.i.d. repetitions is $B = 100$.} \label{fig:CLR_EC}
\end{figure}

\noindent
\textbf{Results and comments.} Figures \ref{fig:RLG_EC}, \ref{fig:RL_EC} and \ref{fig:CLR_EC} show detection rates for cross validation and for the revisited knockoffs method after thresholding with the $W$-threshold and with the gaps-threshold. These detection rates are illustrated on Figures \ref{fig:RLG_EC}, \ref{fig:RL_EC} and \ref{fig:CLR_EC} for respectively linear, logistic and cumulative logit regressions. First, we can note that our procedure is efficient for each of these regressions: the difference of detection rates of the first five covariates and the rest of them is really clear, regardless of the regression coefficients or the choice of the threshold. For linear regression, these two thresholds give similar results whereas for logistic and cumulative logit regressions, gaps-threshold tends to give slightly lower detection rates than $W$-threshold, for both relevant and noisy covariates. \\
In comparison, cross validation provides considerably higher detection rates: although the first five covariates, especially $X_4$ and $X_5$, can be more detected, noisy covariates are also much more detected than with our procedure. For example, for logistic regression in Figure \ref{fig:RL_EC}, noisy covariates are almost all detected less than 20\% with our procedures whereas they are detected between 20\% and 40\% with cross validation. In practice, using cross validation would give more false positive detections than our procedures.\\
Figure \ref{fig:RLG_EC} also show detection rates obtained by Barber and Candes' knockoffs in the linear Gaussian regression model. To perform their procedure, we have to choose a target false discovery rate. In practice, we want it to be small but too small values lead to an infinite threshold and thus an empty estimated model. By default, the FDR is set to $0.1$ but we set it to $0.4$ to avoid too many empty estimated models. For the two different configurations of $\boldsymbol \beta$, we have obtained $4$ empty estimated models on $100$ repetitions. Because of that, detection rates of the noisy covariates tend to be a bit higher than ours. For the same reason, the five first ones are a bit less detected (close to $96$\%, which corresponds to the number of repetitions for which the threshold was not infinite). For $\boldsymbol \beta = (2.5,2,1.5,1,0.5,0,\ldots, 0)$, $X_5$ is yet better detected.\\
All of these three figures illustrate also that detection rates depend on the regression coefficient $\beta$: the higher $\beta$ is, the more the associated covariate is detected. Indeed, for $\boldsymbol \beta = (2.5,2,1.5,1,0.5,0,\ldots,0)$, we can observe that the covariate $X_5$ tends to be less detected than the four first ones. Furthermore, we can notice that some noisy covariates are more detected. For example, this is the case for the noisy covariates $X_{18}, X_{19}, X_{29}, X_{36}, X_{39}$ or $X_{47}$ for $\boldsymbol \beta = (1,1,1,1,1,0,\ldots, 0)$ and for all kind of regressions (since the design matrix is the same). This is probably due to the dependence structure of $\overrightarrow X$. In particular, these covariates are dependent to three of the first five covariates conditionally on the other ones. Similar phenomenon occurs for $\boldsymbol \beta = (2.5,2,1.5,1,0.5,0,\ldots, 0)$ with noisy covariates $X_{19}$, $X_{41}$ and $X_{47}$.\\
Finally, our procedure seems to be quite efficient regardless of the regression model. Nevertheless, results are a little bit better for linear regression. In this case, we remind that %it is more appropriate and powerful to use 
Barber and Candes' procedures \cite{CandesKnockoff,Candes2016knockoff}
%, which moreover 
also provide theoretical guarantees.

\subsubsection{p = 2000}

We present now results for $n = 1000$ observations of $p = 2000$ covariates to illustrate that our procedure is suitable with thousands of covariates. Regression coefficients are set to $\beta_k = \left\{
\begin{array}{l}
  5, $ if $1 \leq k \leq 20, \\
  4, $ if $21 \leq k \leq 40, \\
  3, $ if $41 \leq k \leq 60, \\
  2, $ if $61 \leq k \leq 80, \\
  1, $ if $81 \leq k \leq 100, \\  
  0, $ otherwise.$
\end{array}
\right.$. In the same way as for $p = 50$ (Subsection \ref{p50}), covariates $\textbf{X}$ are the same for each different regression. But they are different according to the regression coefficients $\boldsymbol \beta$. In other words, for a fixed value of $\boldsymbol \beta$, the design matrix $\textbf{X}$ is the same for each of the three regression models. But the response variable $\textbf{Y}$ is simulated according to the regression model and is therefore different. The knockoffs matrix is also different.\\

\begin{figure}[htbp!]
\centering
\subfloat[Revisited knockoffs with $W$-threshold.]{%
\resizebox*{7.11cm}{!}{\includegraphics{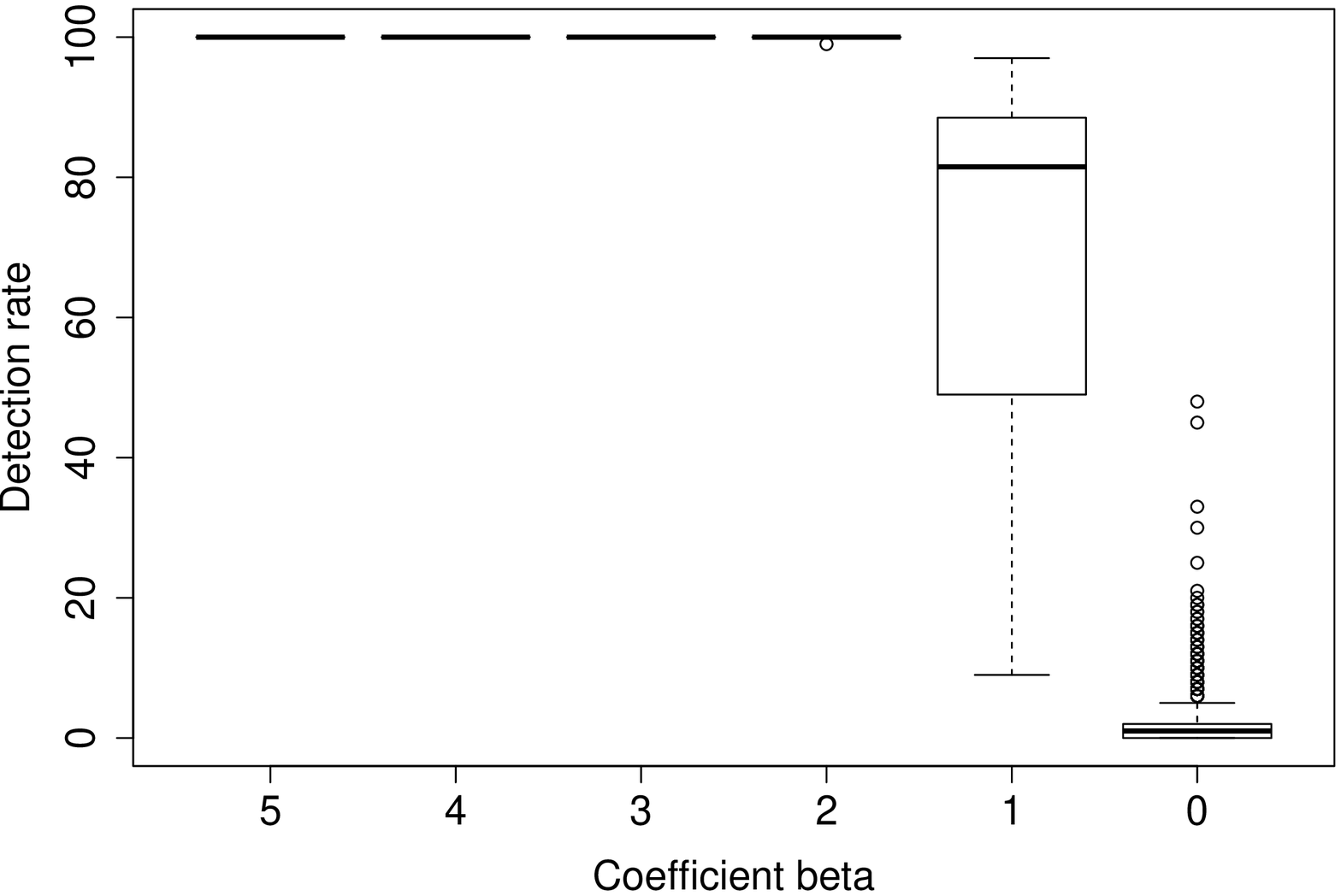}}\label{fig:RLG_2000_W}}\hspace{0pt}
\subfloat[Revisited knockoffs with gaps-threshold.]{%
\resizebox*{7.11cm}{!}{\includegraphics{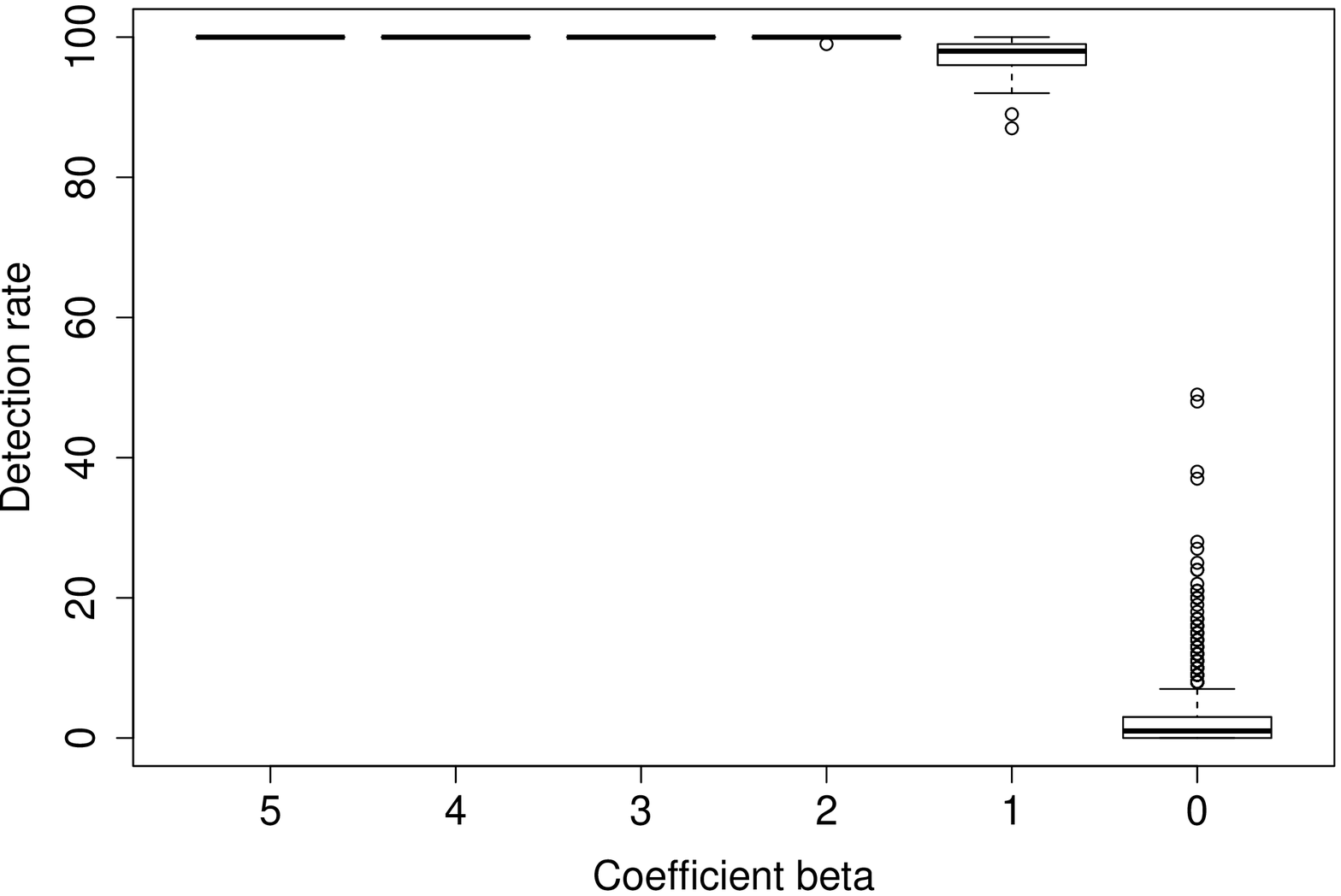}}\label{fig:RLG_2000_gaps}}\vspace{0pt}
\subfloat[Cross validation.]{%
\resizebox*{7.11cm}{!}{\includegraphics{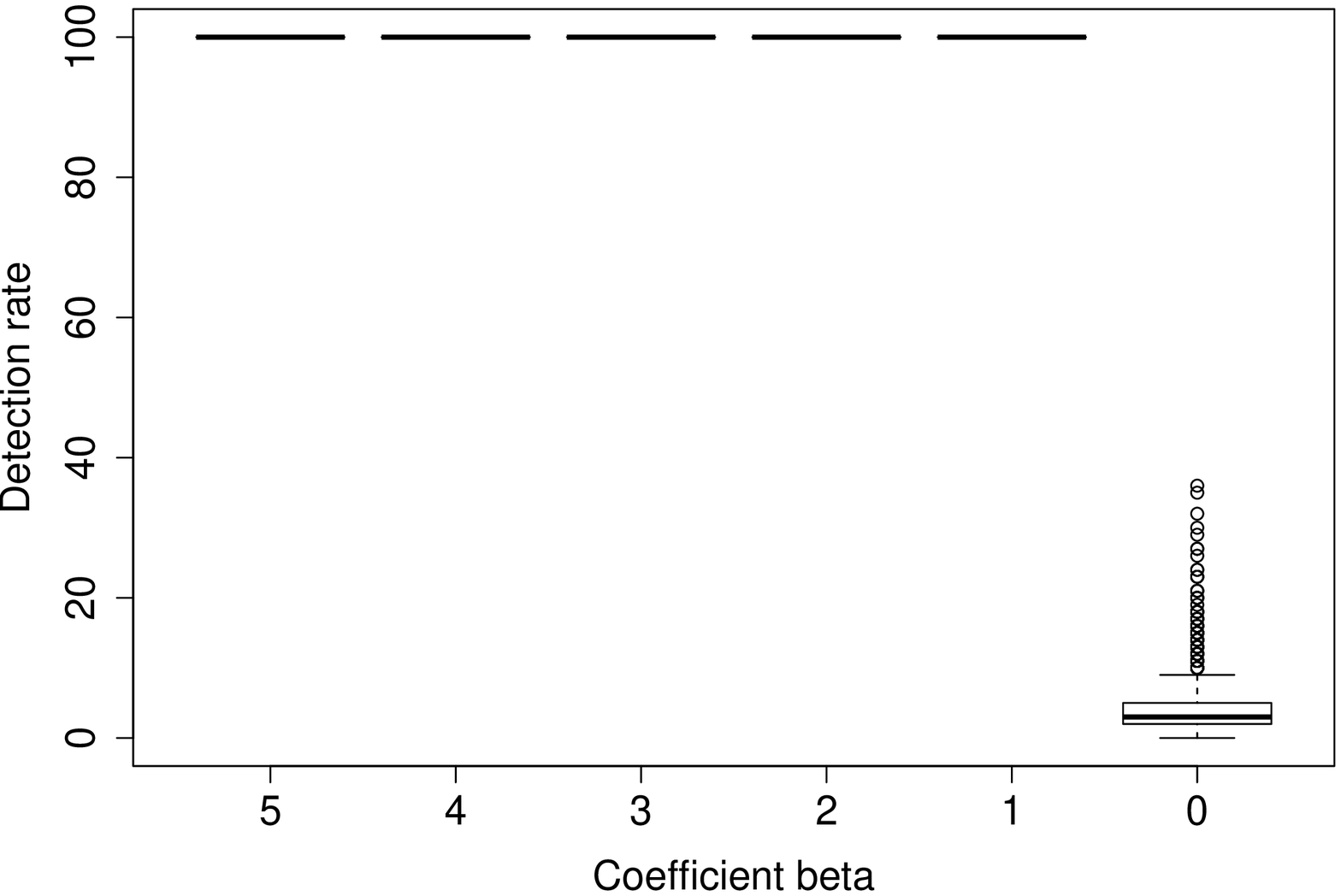}}}\hspace{0pt}
\subfloat[Comparison of detection rates of noisy covariates for the revisited knockoffs method with gaps-threshold and for cross validation.]{%
\resizebox*{7.11cm}{!}{\includegraphics{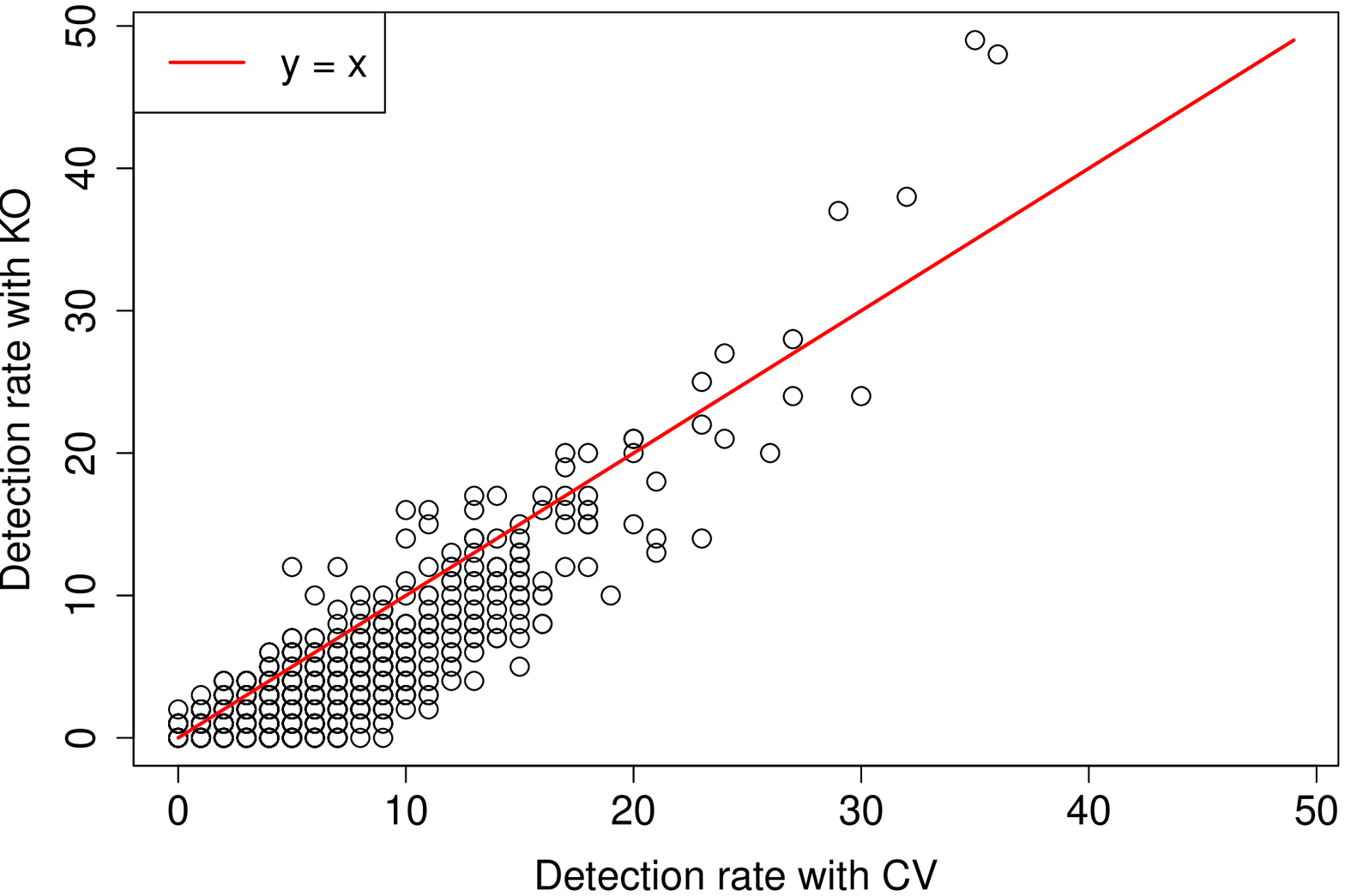}}\label{fig:RLG_2000_NP}}
\caption{Boxplots of detection rates of each covariate according to their regression coefficient $\beta$ for the three methods: revisited knockoffs $W$-threshold (a), gaps-thresholds (b) and cross validation (c). Linear Gaussian regression model with $n = 1000$ observations of $p = 2000$ covariates. Covariates are dependent Gaussian with a random structure. The number of i.i.d. repetitions is $B = 100$.} \label{fig:RLG_2000}
\end{figure}

\begin{figure}[htbp!]
\centering
\subfloat[Revisited knockoffs with $W$-threshold.]{%
\resizebox*{7.11cm}{!}{\includegraphics{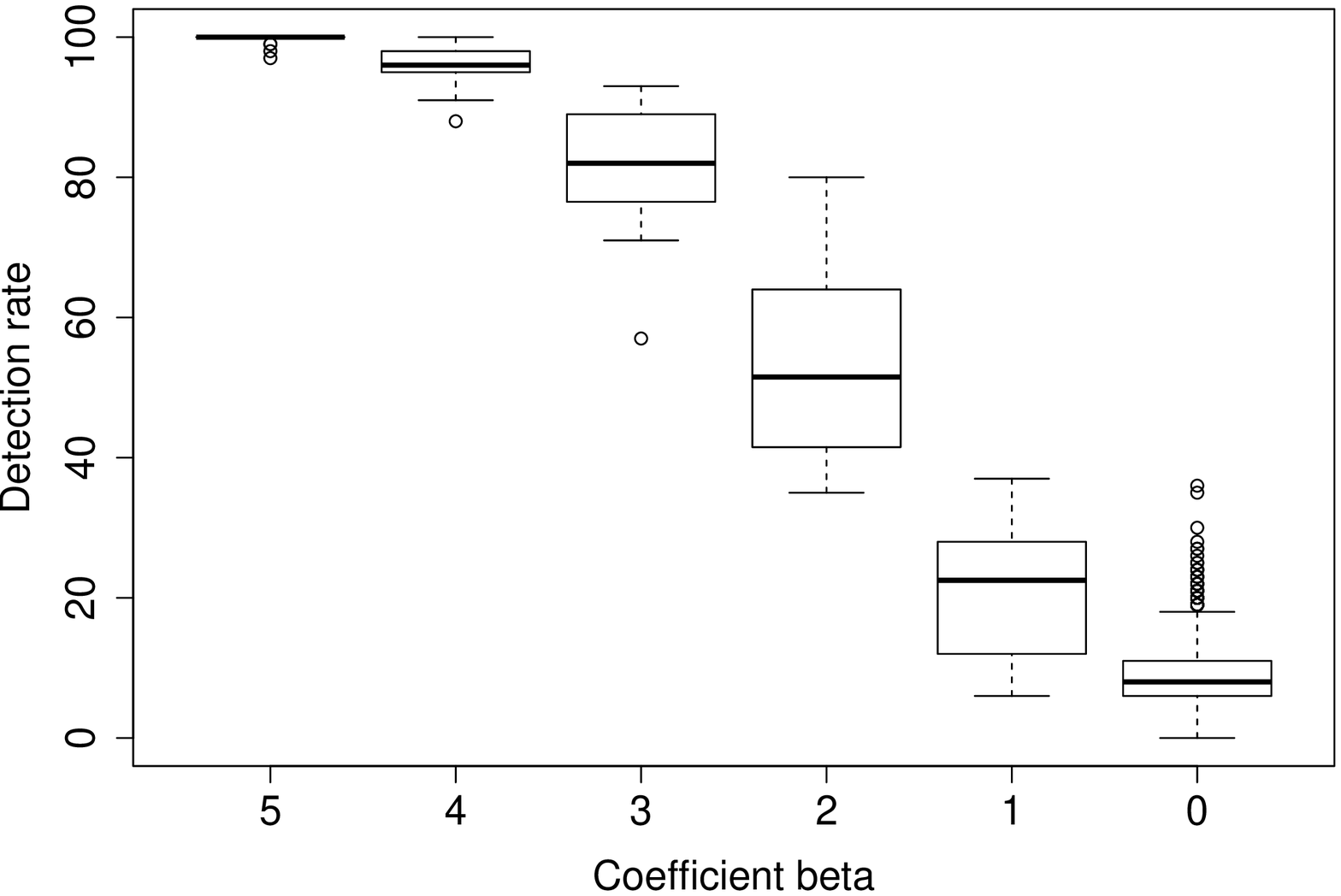}}}\hspace{0pt}
\subfloat[Revisited knockoffs with gaps-threshold..]{%
\resizebox*{7.11cm}{!}{\includegraphics{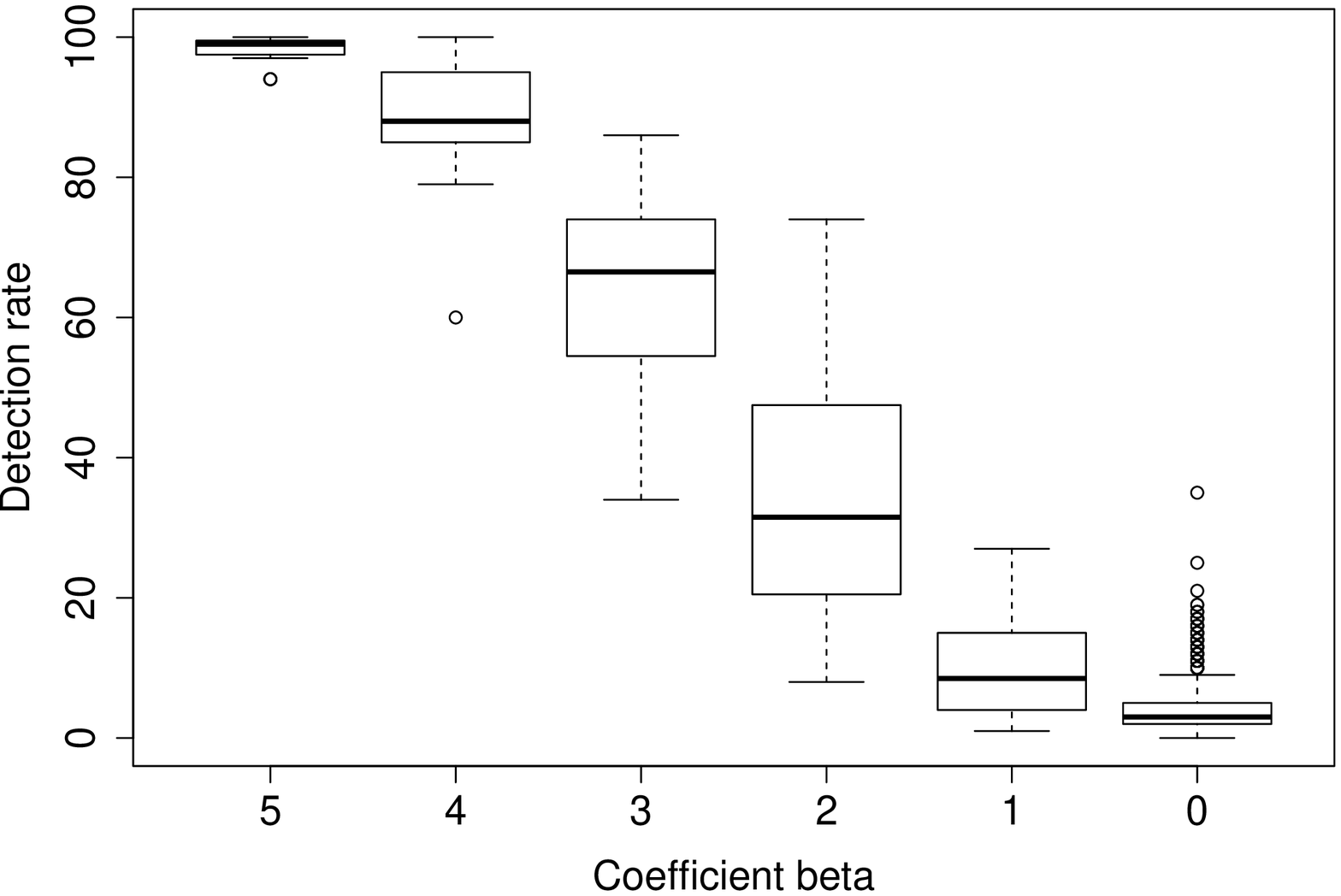}}}\vspace{0pt}
\subfloat[Cross validation.]{%
\resizebox*{7.11cm}{!}{\includegraphics{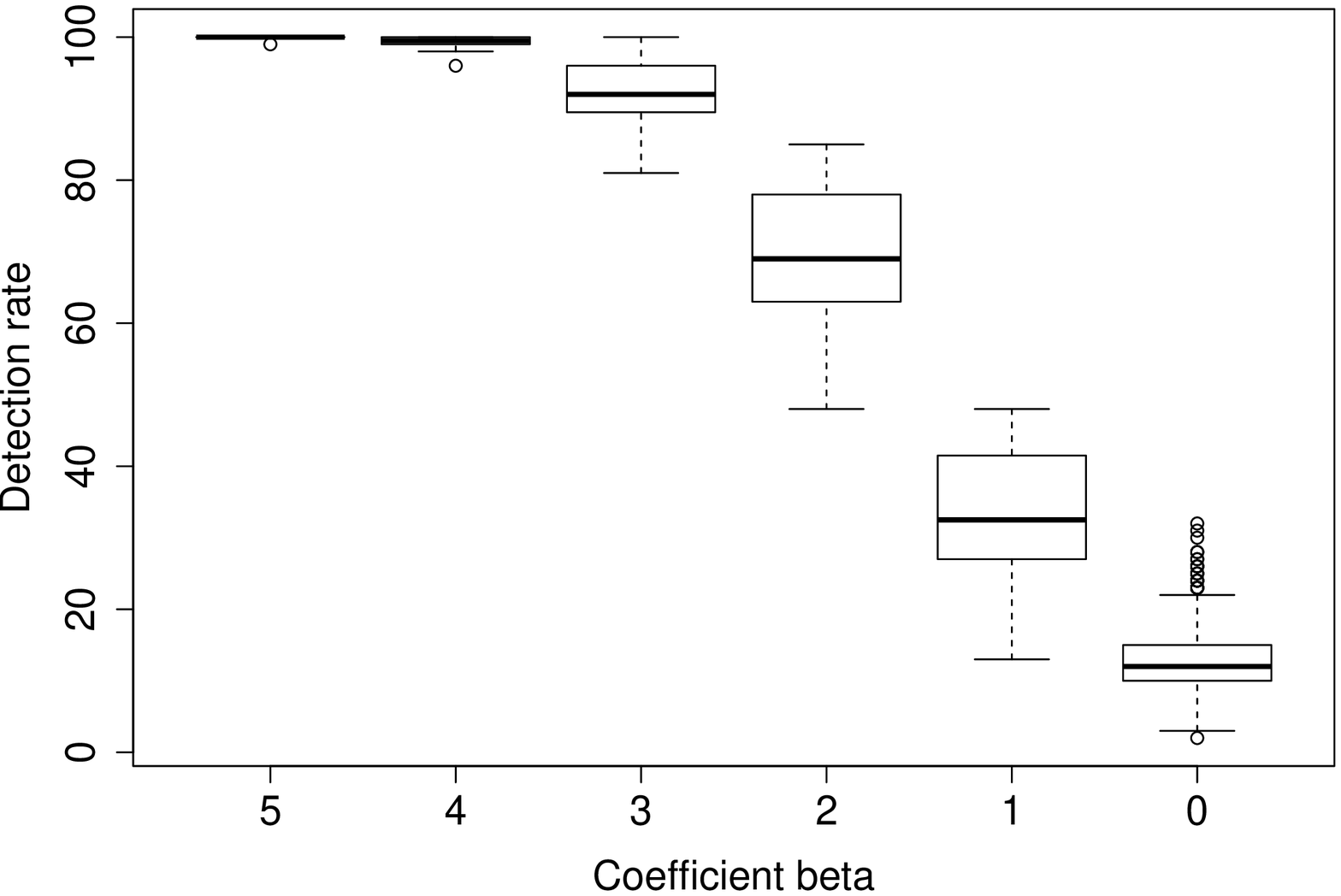}}}\hspace{0pt}
\subfloat[Comparison of detection rates of noisy covariates for the revisited knockoffs method with gaps-threshold and for cross validation.]{%
\resizebox*{7.11cm}{!}{\includegraphics{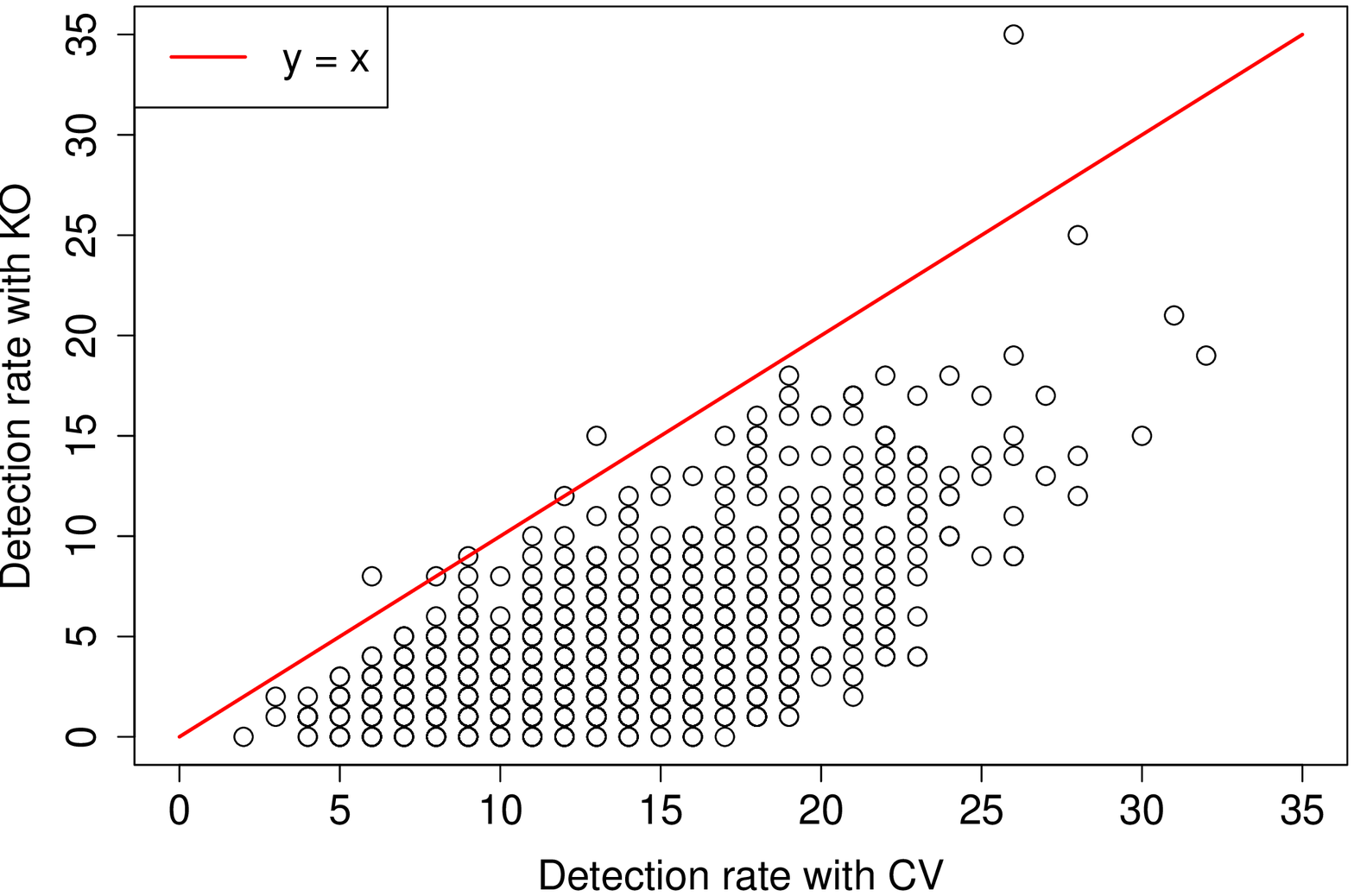}}\label{fig:RL_2000_NP}}
\caption{Boxplots of detection rates of each covariate according to their regression coefficient $\beta$ for the three methods: revisited knockoffs $W$-threshold (a), gaps-thresholds (b) and cross validation (c). Logistic regression model with $n = 1000$ observations of $p = 2000$ covariates. Covariates are dependent Gaussian with a random structure. The number of i.i.d. repetitions is $B = 100$.} \label{fig:RL_2000}
\end{figure}

\begin{figure}[htbp!]
\centering
\subfloat[Revisited knockoffs with $W$-threshold.]{%
\resizebox*{7.11cm}{!}{\includegraphics{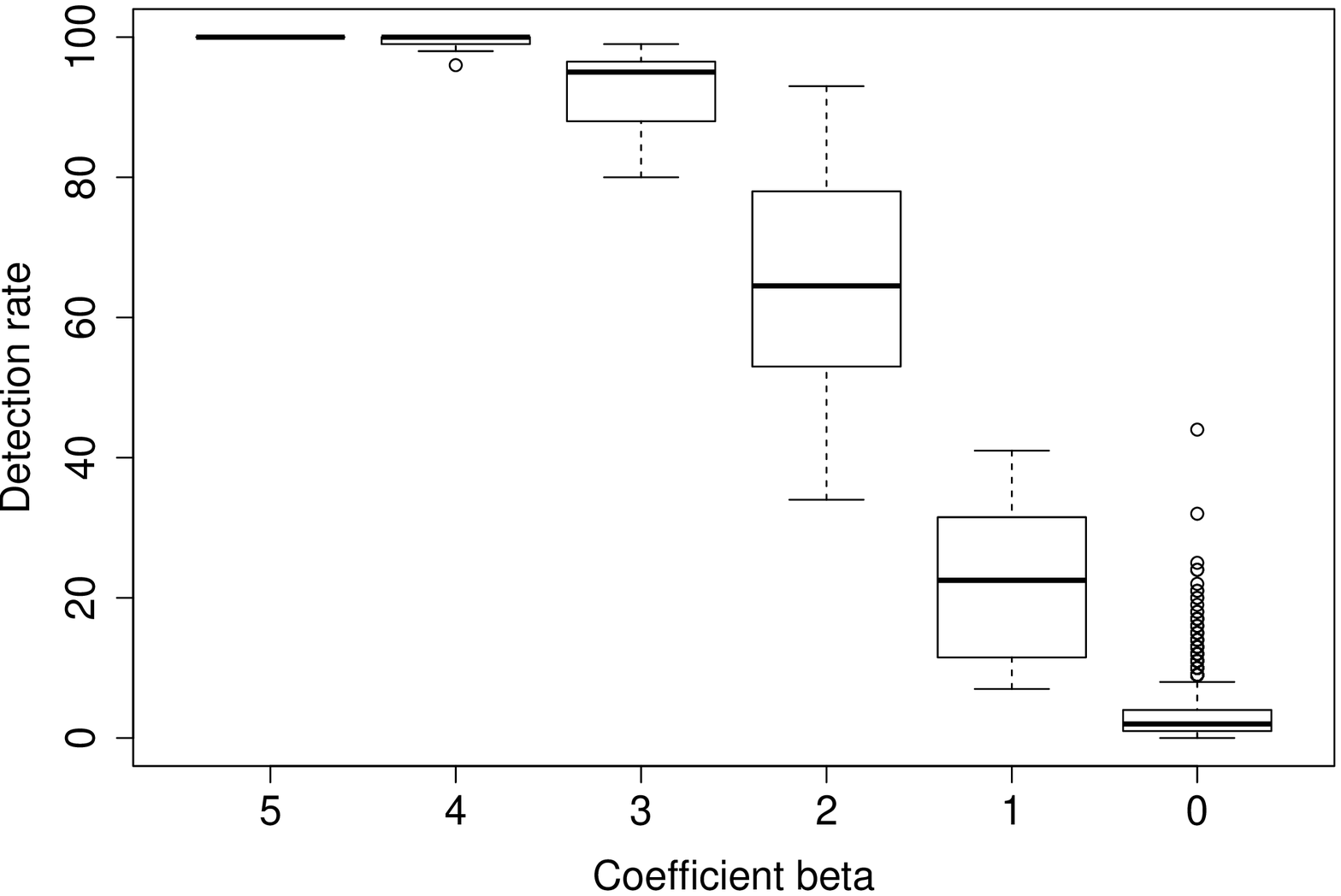}}}\hspace{0pt}
\subfloat[Revisited knockoffs with gaps-threshold..]{%
\resizebox*{7.11cm}{!}{\includegraphics{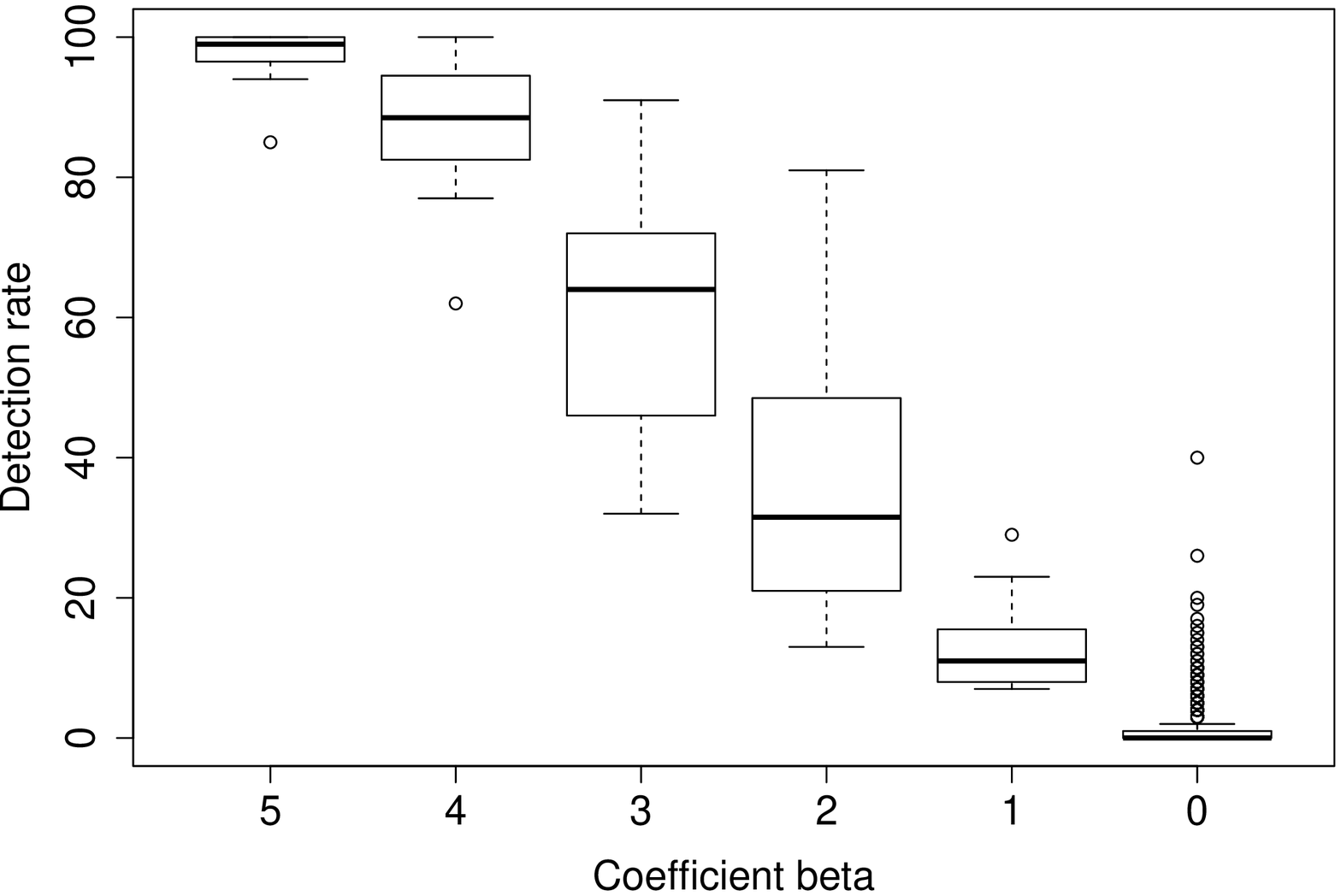}}}\vspace{0pt}
\subfloat[Cross validation.]{%
\resizebox*{7.11cm}{!}{\includegraphics{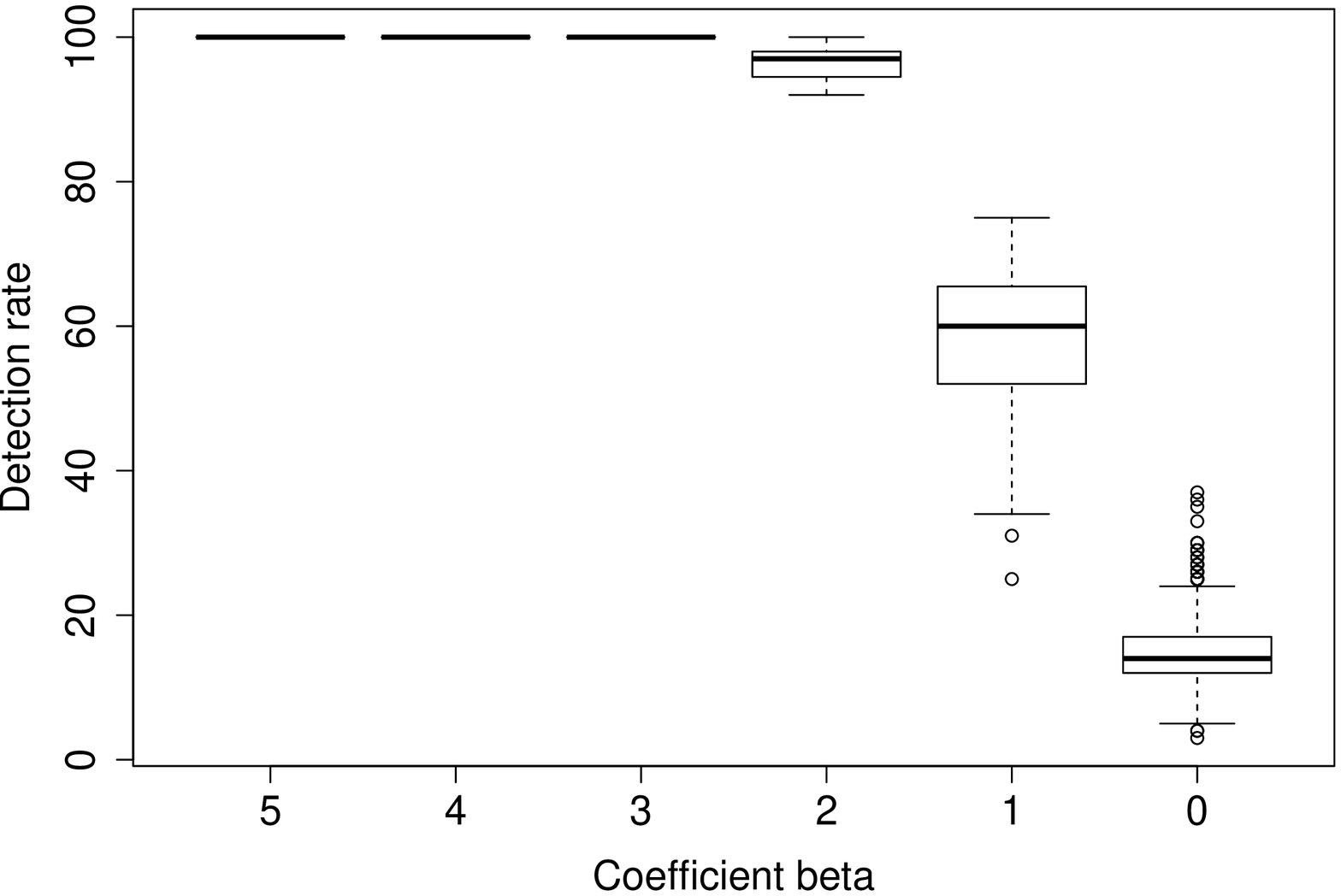}}}\hspace{0pt}
\subfloat[Comparison of detection rates of noisy covariates for the revisited knockoffs method with gaps-threshold and for cross validation.]{%
\resizebox*{7.11cm}{!}{\includegraphics{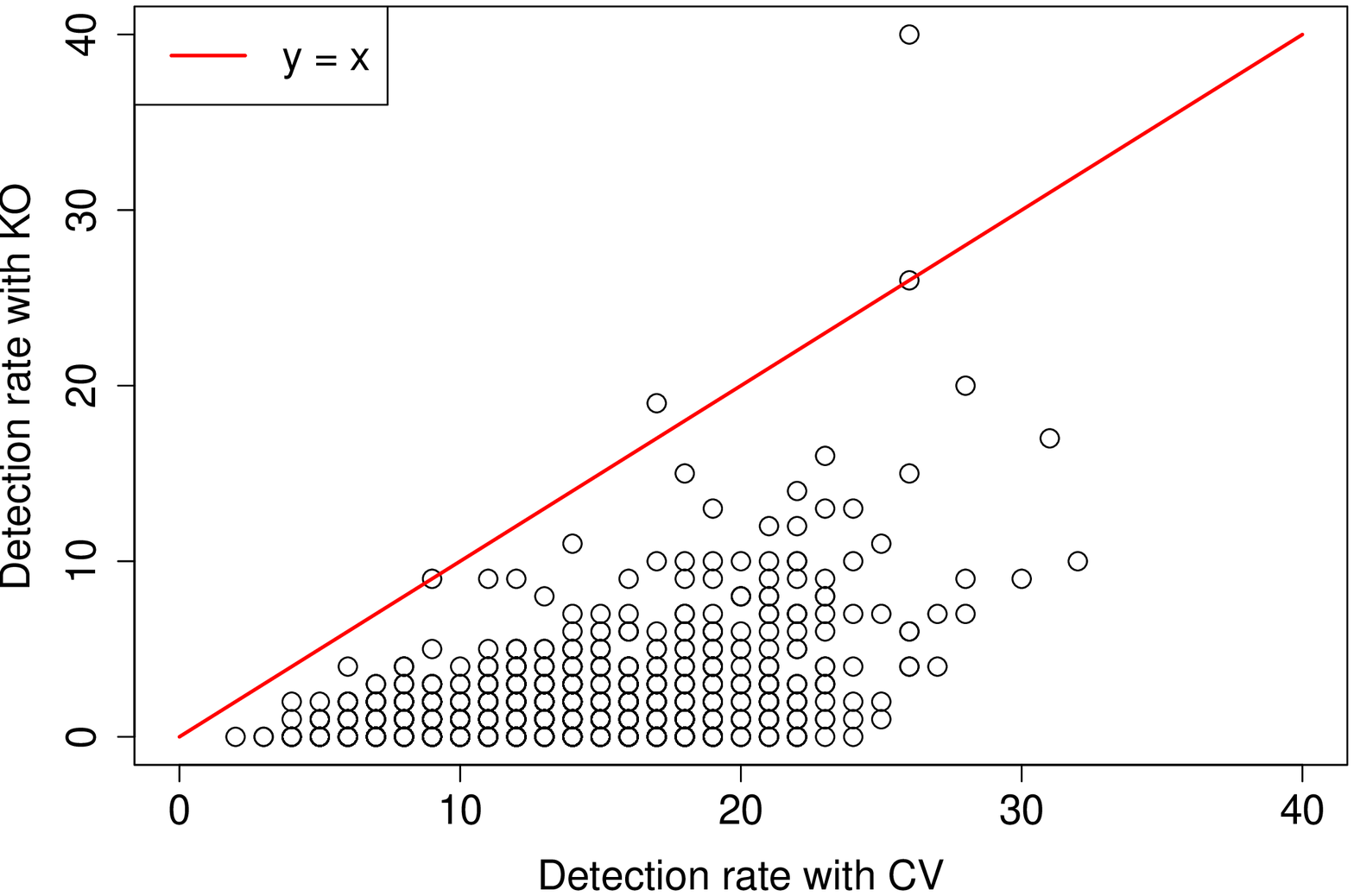}}\label{fig:Cum_2000_NP}}
\caption{Boxplots of detection rates of each covariate according to their regression coefficient $\beta$ for the three methods: revisited knockoffs $W$-threshold (a), gaps-thresholds (b) and cross validation (c). Cumulative logit model with $n = 1000$ observations of $p = 2000$ covariates. Covariates are dependent Gaussian with a random structure. The number of i.i.d. repetitions is $B = 100$.} \label{fig:Cum_2000}
\end{figure}

\noindent
\textbf{Results and comments.} Figures \ref{fig:RLG_2000}, \ref{fig:RL_2000} and \ref{fig:Cum_2000} each contain four graphics: three of them are boxplots of detection rate of each of the 6 groups of covariates according to their regression coefficient $\beta$. These detection rates are respectively obtained with the revisited knockoffs methods and cross validation. In order to compare our method with cross validation for the noisy covariates, we present detection rates of the noisy covariates (these for which $\beta = 0$) obtained with the knockoffs method (with gaps-threshold) in function of detection rates obtained with cross validation in the last graphic.\\
Results in the linear regression framework are presented in Figure \ref{fig:RLG_2000}. Comparing the three boxplots, we can remark that detection rates with revisited knockoffs method with $W$-threshold are lower than with gaps-threshold. More specifically, relevant covariates whose $\beta = 1$ are detected between 10 and 95\% with $W$-threshold whereas they are detected more than 80\% with gaps-threshold. Detection rates with $W$-threshold are lower and more widespread for covariates whose regression coefficient $\beta = 1$ comparing to the other relevant covariates. Cross validation leads to better detection rates for the relevant covariates. However, Figure \ref{fig:RLG_2000_NP} illustrates that most of the noisy covariates have higher detection rates with cross validation than with our procedure. Thus, cross validation gives more false positive detections.\\
Results for logistic and cumulative logit regressions are respectively presented in Figures \ref{fig:RL_2000} and \ref{fig:Cum_2000}. As for $p = 50$, we can note on boxplots that detection rates depend on the regression coefficient $\beta$: for all the three methods, detection rates are decreasing according to $\beta$ that is, the higher $\beta$ is, the more the associated covariates are detected. We also observe this on graphics \ref{fig:RLG_2000_W} and \ref{fig:RLG_2000_gaps} for linear regression, although it is less pronounced. As for linear regression, even if cross validation gives better detection rates for the relevant covariates, it also gives more fase positive detections for the noisy covariates as illustrated in graphics \ref{fig:RL_2000_NP} and \ref{fig:Cum_2000_NP}. This phenomenon is even stronger for these two regression models where almost all of the noisy covariates are more detected with cross validation than with our procedure. Contrary to linear regression, detection rates obtained by the knockoffs method with $W$-threshold are higher than with gaps-threshold and they are higher for both relevant and noisy covariates.\\
% Différence entre les méthodes : plus ou moins adapté et le seuil à choisir peut différer selon la méthode
Although detection rates are better for linear regression, our procedures lead to satisfying results for the three regression models. Even though relevant covariates are not always enough detected, detection rates of noisy covariates are also often very low, especially in comparison with cross validation. On the whole, it seems to be appropriate for sparse models regardless to the regression model and particularly when the goal is to avoid false positive detections. Notice that the threshold ($W$ or gaps-threshold) to be used to avoid false positive detections may vary according to the regression model.

\subsection{Randomness of the procedure} \label{Randomness}

Note that revisited knockoffs procedure and cross validation are both random (which is not the case for Barber and Candes' procedure). Indeed, the former is random in the construction of the knockoffs matrix whereas randomness of cross validation lies in the choice of the folds. Hence, applying several time one of these methods leads to different results. To conclude this section, we compare detection rates obtained by these three methods on the same sample of data. This sample includes $n = 200$ observations of $p = 50$ covariates and consists in the first sample of the $B = 100$ samples used in the Subsection \ref{p50}. Dependence structure of the vector $\overrightarrow X$ is thus also the same as in Subsection \ref{p50}.\\

\begin{figure}[htbp!]
\centering
\subfloat[Linear regression.]{%
\resizebox*{7.11cm}{!}{\includegraphics{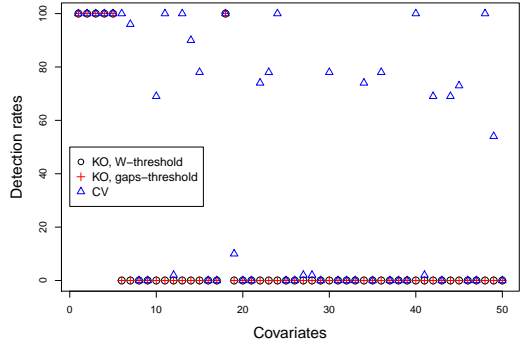}}}\hspace{0pt}
\subfloat[Logistic regression.]{%
\resizebox*{7.11cm}{!}{\includegraphics{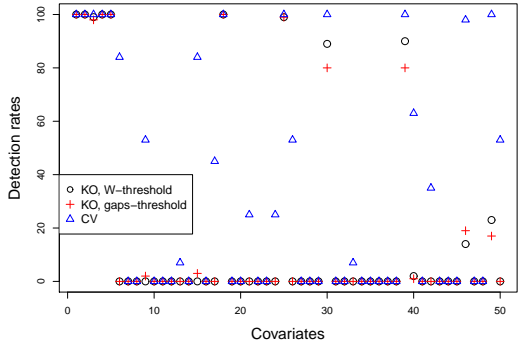}}}
\subfloat[Cumulative logit regression.]{%
\resizebox*{7.11cm}{!}{\includegraphics{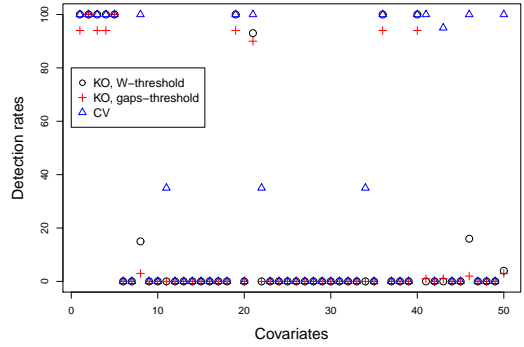}}}
\caption{Detection rates of each covariate for the three methods: revisited knockoffs method with the $W$- and gaps-thresholds (see Subsection \ref{Thresholds}) and cross validation. Detection rates are obtained on 100 repetitions on the same sample of $n = 200$ observations of $p = 50$ covariates with regression coefficients $\boldsymbol \beta = (1,1,1,1,1,0,\ldots, 0)$. Covariates are dependent Gaussian with a random structure.} \label{fig:Random}
\end{figure}

\noindent
\textbf{Results and comments.} Figure \ref{fig:Random} displays detection rates of each covariate using randomness of the three procedures: revisited knockoffs method with $W$ and gaps-thresholds and cross validation. Detection rates are obtained on 100 repetitions on the same sample for the three methods and for the three regression models: linear, logistic and cumulative logit regressions. Regression coefficients are set to $\boldsymbol \beta = (1,1,1,1,1,0,\ldots, 0)$. Thus, only the first five covariates belong to the model. We can notice that these first five covariates are almost always 100\% detected except with gaps-threshold for cumulative logit regression (for which they are still detected more than 95\%). Noisy covariates, that is covariates $X_6, \ldots, X_{50}$, are always less detected by our procedures than by cross validation. However, some of them are wrongly highly detected: $X_{18}$ for linear regression, $X_{18}, X_{25}, X_{30}$ and $X_{39}$ for logistic regression or $X_{19}, X_{21}, X_{36}$ and $X_{40}$ for cumulative logit regression. This is probably due to the sample (in which the dependence structure is more or less pronounced). It should be recalled that this dependence structure is the same as in Subsubsection \ref{p50} for $p = 50$. In comparison with Figures \ref{fig:RLG_EC}, \ref{fig:RL_EC} and \ref{fig:CLR_EC} of Subsubsection \ref{p50}, we can see that these covariates have also higher detection rates. For all these covariates, cross validation gives higher detection rates than our procedures. Cross validation gives also much higher detection rates for some other noisy covariates. For instance, $X_6, X_{11}, X_{13}, X_{24}, X_{40}$ and $X_{48}$ are always detected for linear regression whereas revisited knockoffs with $W$ and gaps threshold never detect them. For cumulative logit model, $X_8, X_{41}, X_{46}$ and $X_{50}$ are always detected whereas our procedures detect them less than 15\%.\\

In practice, with real data, this randomness opens up to further ways to perform variable selection.

%---------------------------------------------------------------------------------------

% discussion

\section{Discussion}

In this paper, we proposed a method for variable selection in regression models based on the construction of a matrix of knockoffs of the covariates. This method is quite intuitive and suitable for many types of regressions, including when the number of observations is much smaller than the number of covariates. Two different thresholds can be chosen, leading to two procedures, which have been implementend in the R package \texttt{kosel}. We have seen that these procedures both turn out to be very pertinent and efficient as the many and diverse simulations exemplify. Our two procedures are particularly appropriate when the goal is to avoid false positive detections. Indeed, even if there are false negative detections, there are also a very small rate of false positive detections. Simulations show also that efficiency of our procedures depends on the regression model. In general, we can try the two thresholds and choose results according to its target. Furthermore, randomness of our procedures provides other techniques to perform variable selection. \\
Nonetheless, in the case of linear Gaussian regression, Barber and Candes' procedures also offer theoretical guarantees. \\
In addition, our procedures give better results than cross validation with regard to false positive detections, even when we make use of randomness. However, if we aim at overselecting, it is more appropriate to use other techniques such as cross validation. \\

%---------------------------------------------------------------------------------------

\bibliographystyle{plain}
\bibliography{Biblio}

\end{document}